\RequirePackage{snapshot}
\documentclass[nospthms]{svjour3}                     %
\def\makeheadbox{}  %

\usepackage{microtype}
\usepackage[colorlinks,citecolor={blue},urlcolor={blue},linkcolor={blue}]{hyperref}
\usepackage{fix-cm}
\usepackage{amsmath}
\usepackage{amssymb}
\usepackage{graphicx}
\usepackage{subfig}
\usepackage{cite}
\usepackage{braket}
\numberwithin{equation}{section}

\usepackage{amsthm}
\newtheorem{theorem}{Theorem}[section]

\newtheorem{lemma}[theorem]{Lemma}

\newtheorem{proposition}[theorem]{Proposition}

\input{macros}
\usepackage{boxedminipage}
\usepackage{color}
\definecolor{softblue}{rgb}{0.90,0.92,1.00}
\makeatletter\newenvironment{btheorem}{%
\begin{lrbox}{\@tempboxa}\begin{minipage}{0.97\textwidth}\begin{theorem}}%
{\end{theorem}\end{minipage}\end{lrbox}%
\par\hbox{}\noindent%
{\setlength{\fboxsep}{0pt}\colorbox{softblue}{\setlength{\fboxsep}{4pt}\begin{boxedminipage}{\textwidth}\usebox{\@tempboxa}\end{boxedminipage}}}%
\vspace{0.5\baselineskip}}
\makeatother
\makeatletter%
{\end{lemma}\end{minipage}\end{lrbox}%
\par\hbox{}\noindent%
{\setlength{\fboxsep}{0pt}\colorbox{softblue}{\setlength{\fboxsep}{4pt}\begin{boxedminipage}{\textwidth}\usebox{\@tempboxa}\end{boxedminipage}}}%
\vspace{0.5\baselineskip}}
\makeatother
\makeatletter\newenvironment{bproposition}{%
\begin{lrbox}{\@tempboxa}\begin{minipage}{0.97\textwidth}\begin{proposition}}%
{\end{proposition}\end{minipage}\end{lrbox}%
\par\hbox{}\noindent%
{\setlength{\fboxsep}{0pt}\colorbox{softblue}{\setlength{\fboxsep}{4pt}\begin{boxedminipage}{\textwidth}\usebox{\@tempboxa}\end{boxedminipage}}}%
\vspace{0.5\baselineskip}}
\makeatother
\def\expval{\mathbb{E}}
\def\var{\mathbb{V}}
\def\sample{\Sscr}

\def\phis{\phi_{\scriptscriptstyle\sample}}

\title{Robust inversion, dimensionality reduction, \\and randomized sampling%
    \thanks{This work was in part financially supported by the
    Natural Sciences and Engineering Research Council of Canada
    Discovery Grant (22R81254) and the Collaborative Research and
    Development Grant DNOISE II (375142-08). %
    This research was carried
    out as part of the SINBAD II project with support from the
    following organizations: BG Group, BPG, BP, Chevron, Conoco
    Phillips, Petrobras, PGS, Total SA, and WesternGeco.%
  }
}

\author{%
  Aleksandr Aravkin
  \\\hbox{Michael P. Friedlander}
  \\\hbox{Felix J. Herrmann}
  \\\hbox{Tristan van Leeuwen}}

\authorrunning{A.\@ Aravkin, M. P. Friedlander, F. J. Herrmann, and
  T. van Leeuwen} %

\institute{%
              A. Aravkin, F. J. Herrmann, and T. van Leeuwen \at
              Dept. of Earth and Ocean Sciences, %
              University of British Columbia, %
              Vancouver, BC, Canada %
            \\\email{\{saravkin,fherrmann,tleeuwen\}@eos.ubc.ca}
          \and
              M. P. Friedlander \at %
              Dept. of Computer Science, %
              University of British Columbia, %
              Vancouver, BC, Canada %
            \\\email{mpf@cs.ubc.ca}
}

\date{March 1, 2012}

\begin{document}

\maketitle

\begin{abstract}
  We consider a class of inverse problems in which the forward model
  is the solution operator to linear ODEs or PDEs. This class admits
  several di\-men\-sionality-reduction techniques based on data averaging
  or sampling, which are especially useful for large-scale
  problems. We survey these approaches and their connection to
  stochastic optimization. The data-averaging approach is only viable,
  however, for a least-squares misfit, which is sensitive to outliers
  in the data and artifacts unexplained by the forward model. This
  motivates us to propose a robust formulation based on the Student's
  t-distribution of the error. We demonstrate how the corresponding
  penalty function, together with the sampling approach, can obtain
  good results for a large-scale seismic inverse problem with 50\%
  corrupted data.

  \keywords{inverse problems \and seismic inversion
    \and stochastic optimization \and robust estimation}
\end{abstract}

\section{Introduction}
\label{sec:intro}

Consider the generic parameter-estimation scheme in which we conduct $m$
experiments, recording the corresponding experimental input vectors
$\{q_1,q_2,\ldots,q_m\}$ and observation vectors $\{d_1,d_2,\ldots,d_m\}$.
We model the data for given parameters $x\in\Real^n$ by
\begin{equation}
  \label{eq:Fi}
  d_i = F_i(x)q_i + \epsilon_i \text{for} i=1,\ldots,m,
\end{equation}
where observation $d_i$ is obtained by the linear action of the
forward model $F_i(x)$ on known source parameters $q_i$, and
independent errors $\epsilon_i$ capture
the discrepancy between $d_i$ and prediction
$F_i(x)q_i$. The class of models captured by this representation
includes solution operators to any linear (partial) differential
equation with boundary conditions, where the $q_i$ are the
right-hand sides of the equations.  A special case arises when $F_i
\equiv F$, i.e., the forward model is the same for each experiment.

Inverse problems based on these forward models arise in a variety of
applications, including medical imaging and seismic exploration, in
which the parameters $x$ usually represent particular physical
properties of a material. We are particularly motivated by the
full-waveform inversion (FWI) application in seismology, which is used
to image the earth's subsurface~\cite{Tarantola1984}.  In
full-waveform inversion, the forward model $F$ is the solution
operator of the wave equation composed with a restriction of the full
solution to the observation points (receivers); $x$ represents
sound-velocity parameters for a (spatial) 2- or 3-dimensional mesh;
the vectors $q_i$ encode the location and signature of the $i$th
source experiment; and the vectors $d_i$ contain the corresponding
measurements at each receiver.  A typical survey in exploration
seismology may contain thousands of experiments (shots), and global
seismology relies on natural experiments provided by measuring
thousands of earthquakes detected at seismic stations around the
world.  Standard data-fitting algorithms may require months of CPU
time on large computing clusters to process this volume of data and
yield coherent geological information.

Inverse problems based on the forward models that
satisfy~\eqref{eq:Fi} are typically solved by minimizing some measure
of misfit, and have the general form
\begin{equation}
  \label{eq:5}
  \minimize{x}\quad \phi(x) := \frac{1}{m}\sum_{i=1}^m \phi_i(x),
\end{equation}
where each $\phi_i(x)$ is some measure of the residual
\begin{equation} \label{eq:2}
 r_i(x) := d_i - F_i(x)q_i
\end{equation}
between the observation and prediction of the $i$th experiment.  The
classical approach is based on the least-squares penalty
\begin{equation}
  \label{eq:15}
  \phi_i(x) = \norm{ r_i(x) }^2.
\end{equation}
This choice can be interpreted as finding the maximum
  likelihood (ML) estimate of $x$, given the
assumptions that the errors $\epsilon_i$ are independent and follow a
Gaussian distribution.

Formulation~\eqref{eq:5} is general enough to capture a variety of models,
including many familiar examples. If the $d_i$ and $q_i$ are
scalars, and the forward model is linear, then standard least-squares
\[
  \phi_i(x) = \half(a_i\T x - d_i)^2
\]
easily fits into our general formulation.
More generally, ML estimation is based on the
form
\begin{equation*}
 \phi_i(x) = -\log p_i\big(r_i(x)\big),
\end{equation*}
where $p_i$ is a particular probability density function of $\epsilon_i$.

\subsection{Dimensionality reduction} \label{sec:dr}

Full-waveform inversion is a prime example of an application in which
the cost of evaluating each element in the sum of $\phi$ is very
costly: every residual vector $r_i(x)$---required to evaluate one
element in the sum of \eqref{eq:5}---entails solving a partial
differential equation on a 2D or 3D mesh with thousands of grid points
in each dimension.  The scale of such problems is a motivation for
using dimensionality reduction techniques that address small portions
of the data at a time.

The least-squares objective~\eqref{eq:15} allows for a powerful form
of data aggregation that is based on randomly fusing groups of
experiments into ``meta'' experiments, with the effect of
reducing the overall problem size. The aggregation scheme is based on Haber et
al.'s~\cite{haber10} observation that for this choice of penalty, the
objective is connected to the trace of a residual matrix. That is, we can
represent the objective of~\eqref{eq:5} by
\begin{equation}\label{eq:17}
  \phi(x) = \frac1m\sum_{i=1}^m\norm{r_i(x)}^2
            \equiv \frac{1}{m}\trace\big(R(x)^T R(x)\big)\\,
\end{equation}
where
\[
  R(x) := [r_1(x), r_2(x), \ldots, r_m(x) ]
\]
collects the residual vectors~\eqref{eq:2}. Now consider a small
sample of $s$ weighted averages of the data, i.e.,
\[
 \dtilde_j = \sum_{i=1}^m w_{ij} d_i
 \text{and}
 \qtilde_j = \sum_{i=1}^m w_{ij} q_i,
 \quad
 j = 1,\ldots,s,
\]
where $s\ll m$ and $w_{ij}$ are random variables, and collect the
corresponding $s$ residuals $\rtilde_j(x)=\dtilde_j - F_j(x)\qtilde_j$
into the matrix $R\W(x) := [\rtilde_1(x), \rtilde_2(x), \ldots,
\rtilde_s(x) ]$. Because the residuals are linear in the data, we can
write compactly
\[
 R\W(x) := R(x)W
 \text{where}
 W := (w_{ij}).
\]
Thus, we may consider the sample function
\begin{equation} \label{eq:3}
  \phi\W(x) = \frac1s\sum_{j=1}^s\norm{\rtilde_j(x)}^2
            \equiv \frac{1}{s}\trace\big( R\W(x)\T R\W(x) \big)
\end{equation}
based on the $s$ averaged residuals.
Proposition~\ref{prop:generalSampling} then follows directly from
Hutchinson's~\cite[\S2]{Hutchinson:1990} work on stochastic trace
estimation.
\begin{bproposition}
\label{prop:generalSampling}
If $\expval[WW^T] = I$, then
\begin{equation*}
 \expval\big[\phi\W(x)\big] = \phi(x)
 \text{and}
 \expval[\nabla\phi\W(x)] = \nabla\phi(x).
\end{equation*}
\end{bproposition}

Hutchinson proves that if the weights $w_{ij}$ are drawn independently
from a Rademacher distribution, which takes the values $\pm1$ with
equal probability, then the stochastic-trace estimate has minimum
variance. Avron and Toledo~\cite{AvronToledo:2011} compare the quality
of stochastic estimators obtained from other distributions. Golub and
von Matt~\cite{GoluMatt:1991} report the surprising result that the
estimate obtained with even a single sample ($s=1$) is often of high
quality. Experiments that use the approach in FWI give evidence that
good estimates of the true parameters can be obtained at a fraction of
the computational cost required by the full
approach~\cite{Krebs09,leeuwen2011,HFY:2011}.

\subsection{Approach}

Although the least-squares approach enjoys widespread use, and
naturally accommodates the dimensionality-reduction technique just
described, it is known to be unsuitable for non-Gaussian errors,
especially for cases with very noisy or corrupted data often encountered in practice.
The least-squares formulation also breaks down in the face of systematic features of the data that
are unexplained by the model $F_i$.

Our aim is to characterize the benefits of robust inversion and to
describe randomized sampling schemes and optimization algorithms
suitable for large-scale applications in which even a single evaluation
of the forward model and its action on $q_i$ is computationally
expensive. (In practice, the product $F_i(x)q_i$ is evaluated as a
single unit.) We interpret these sampling schemes, which include
the well-known incremental-gradient
algorithm~\cite{nedic2000convergence}, as dimensionality-reduction
techniques, because they allow algorithms to make progress using only
a portion of the data.

This paper is organized into the following components:

{\it Robust statistics} (\S\ref{sec:mle}). We survey robust approaches
from a statistical perspective, and present a robust approach based on
the heavy-tailed Student's t-distribution.  We show that all
log-concave error models share statistical properties that
differentiate them from heavy-tailed densities (such as the Student's
t) and limit their ability to work in regimes with large outliers or
significant systematic corruption of the data.  We demonstrate that
densities outside the log-concave family allow extremely robust
formulations that yield reasonable inversion results even in the face
of major data contamination.

{\it Sample average approximations} (\S\ref{sec:sampling}). We propose
a dimensionality-reduction technique based on sampling the available
data, and characterize the statistical properties that make it
suitable as the basis for an optimization algorithm to solve the
general inversion problem~\eqref{eq:5}. These techniques can be used
for the general robust formulation described in \S\ref{sec:mle}, and
for formulations in which forward models $F_i$ vary with $i$.

{\it Stochastic optimization} (\S\ref{sec:semistochastic}) We review
stochastic-gradient, randomized in\-cremental-gradient, and
sample-average methods. We show how the assumptions required by each
method fit with the class of inverse problems of interest, and can be
satisfied by the sampling schemes discussed in \S\ref{sec:sampling}.

{\it Seismic inversion} (\S\ref{sec:full-wavef-invers}) We test the
proposed sample-average approach on the robust formulation of the FWI
problem. We compare the inversion results obtained with the new
heavy-tailed approach to those obtained using robust log-concave
models and conventional methods, and demonstrate that a useful
synthetic velocity model can be recovered by the heavy-tailed robust
method in an extreme case with 50\% missing data.  We also compare the
performance of stochastic algorithms and deterministic approaches, and
show that the robust result can be obtained using only 30\% of the
effort required by a deterministic approach.

\section{Robust Statistics}
\label{sec:mle}

A popular approach in robust regression is to replace the
least-squares penalty~\eqref{eq:15} on the residual with a penalty
that increases more slowly than the 2-norm. (Virieux and
Operto~\cite{VirieuxOperto2009} discuss the difficulties with
least-squares regression, which are especially egregious in seismic
inversion.)

One way to derive a robust approach of this form is to assume that the
noise $\epsilon_i$ comes from a particular non-Gaussian probability
density, $p_i$, and then find the maximum likelihood (ML)
estimate of the parameters $x$ that maximizes the
likelihood that the residual vectors $r_i(x)$ are realizations of the
random variable $\epsilon_i$, given the observations $d_i$. Because
the negative logarithm is monotone decreasing, it is natural to
minimize the negative log of the likelihood function rather than
maximizing the likelihood itself. In fact, when the distribution of
the errors $\epsilon_i$ is modeled using a log-concave density \[p(r)
\propto \exp\big(-\rho(r)\big),\] with a convex loss function $\rho$,
the ML estimation problem is equivalent to the
formulation~\eqref{eq:5}, with
\begin{equation}\label{eq:18}
  \phi_i(x) = \rho(r_i(x))
  \text{for}
  i=1,\ldots,m.
\end{equation}

One could also simply start with a penalty $\rho$ on $r_i(x)$, without
explicitly modelling the noise density; estimates obtained this way
are generally known as M-estimates~\cite{Huber:1981}.  A popular
choice that follows this approach is the Huber
penalty~\cite{Huber:1981,HubRon,Mar}.

Robust formulations are typically based on convex penalties $\rho$---or
equivalently, on log-concave densities for $\epsilon_i$---%
that look quadratic near $0$ and increase linearly far from zero.
Robust penalties, including the 1-norm and Huber, for electromagnetic
inverse problems are discussed by Farquaharson and Oldenburg
in~\cite{GJI:GJI555}.  Guitton and Symes~\cite{Symes2003} consider the
Huber penalty in the seismic context, and they cite many previous
examples of the use of 1-norm penalty in geophysics. Huber and 1-norm
penalties are further compared on large-scale seismic problems by
Brossier et al.~\cite{Brossier2010}, and a Huber-like (but strictly
convex) hyperbolic penalty is described by Bube and
Nemeth~\cite{Bube2007}, with the aim of avoiding possible
non-uniqueness associated with the Huber penalty.

Clearly, practitioners have a preference for convex formulations.
However, it is important to note that
\begin{itemize}
\item for nonlinear forward models $F_i$, the optimization
  problem~\eqref{eq:5} is typically nonconvex even for convex
  penalties $\rho$ (it is difficult to satisfy the compositional
  requirements for convexity in that case);
\item even for linear forward models $F_i$, it may be beneficial to
  choose a nonconvex penalty in order to guard against
  outliers in the data.
\end{itemize}
We will justify the second point from a statistical
perspective. Before we proceed with the argument, we introduce the
Student's t-density, which we use in designing our robust method for
FWI.

\subsection{Heavy-tailed distribution: Student's t}
Robust formulations using the Student's t-distribution have been shown
to outperform log-concave formulations in various
applications~\cite{AravkinThesis2010}.  In this section, we introduce
the Student's t-density, explain its properties, and establish a
result that underscores how different heavy-tailed distributions are
from those in the log-concave family.

\begin{figure}[t]
  \centering
  \begin{tabular}{@{}ccc@{}}
  \includegraphics[width=.3\linewidth]{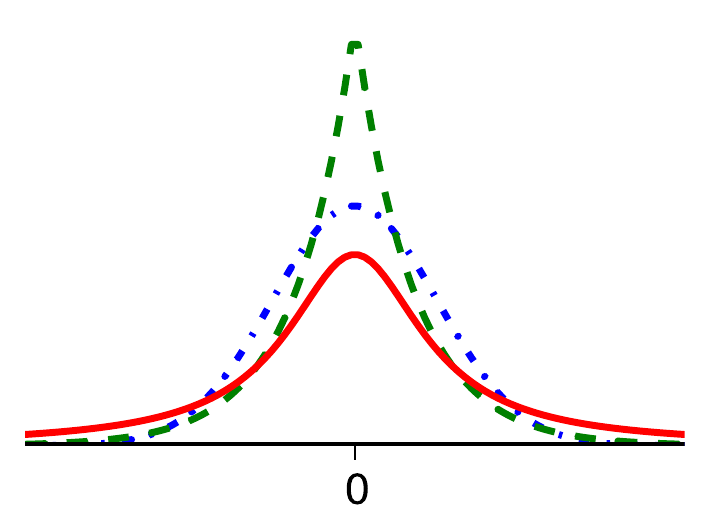}&
  \includegraphics[width=.3\linewidth]{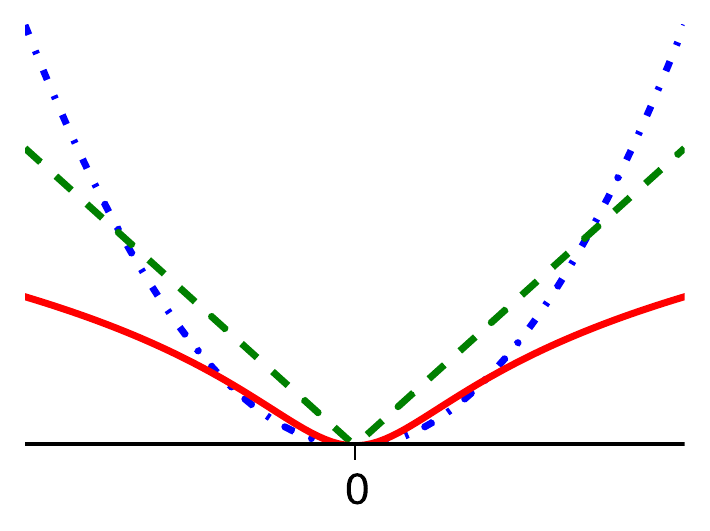}&
  \includegraphics[width=.3\linewidth]{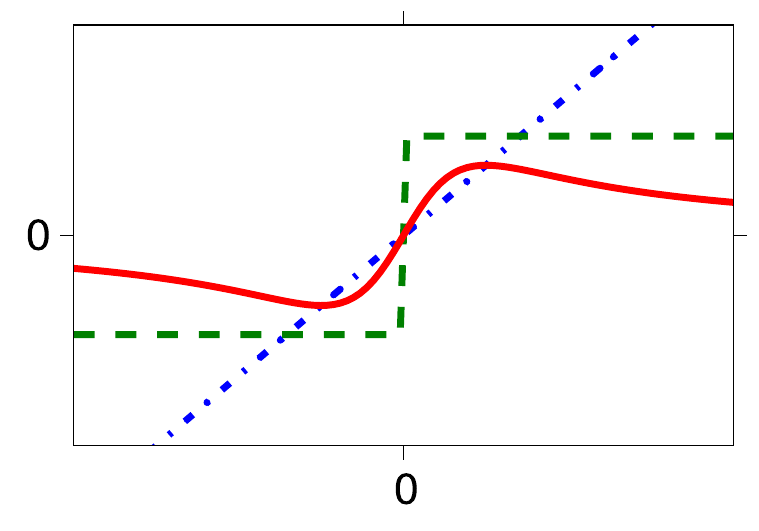}
  \\{\small (a)}&{\small (b)}&{\small (c)}
  \end{tabular}
  \caption{The Gaussian ({\small$\cdot-$}), Laplace ({\small$--$}), and Student's t-
    ({\small ---}) distributions: (a) densities, (b) penalties, and (c)
    influence functions.}
  \label{fig:distributions}
\end{figure}

The scalar Student's t-density function with mean $\mu$ and
positive degrees-of-freedom parameter $\nu$ is given by
\begin{equation}
  \label{eq:st_pdf}
  p(\,r \mid \mu, \nu\,)
  \propto \big(1 + (r - \mu)^2/\nu \big)^{-(1 + \nu)/2}.
\end{equation}
The density is depicted in Figure~\ref{fig:distributions}(a).
The parameter $\nu$ can be understood by recalling the origins of the
Student's t-distribution.  Given $n$ i.i.d. Gaussian variables $x_i$
with mean $\mu$, the normalized sample mean
\begin{equation}
  \label{StudentChar}
  \frac{ \bar x - \mu}{S/\sqrt{n}}
\end{equation}
follows the Student's t-distribution with $\nu = n-1$, where the
sample variance $S^2=\frac{1}{n-1}\sum (x_i - \bar x)^2$ is
distributed as a $\chi^2$ random variable with $n-1$ degrees of
freedom.  As $\nu \rightarrow \infty$, the
characterization~\eqref{StudentChar} immediately implies that the
Student's t-density converges pointwise to the density of
$N(0,1)$. Thus, $\nu$ can be interpreted as a tuning parameter: for
low values one expects a high degree of non-normality, but as $\nu$
increases, the distribution behaves more like a Gaussian
distribution. This interpretation is highlighted
in~\cite{Lange1989}. %

For a zero-mean Student's t-distribution ($\mu=0$), the log-likelihood
of the density~\eqref{eq:st_pdf} gives rise to the nonconvex penalty
function
\begin{equation} \label{eq:19}
  \rho(r) = \log(1 + r^2 / \nu),
\end{equation}
which is depicted in Figure~\ref{fig:distributions}(b).
The nonconvexity of this penalty is equivalent to the sub-exponential
decrease of the tail of the Student's t-distribution, which goes to $0$
at the rate $1/r^{\nu + 1}$ as $r\rightarrow \infty$.

The significance of these so-called \emph{heavy tails} in outlier
removal becomes clear when we consider the following question:
Given that a scalar residual deviates from the mean by more than $t$,
what is the probability that it actually deviates by more than $2t$?

The 1-norm is the slowest-growing convex penalty, and is induced by
the Laplace distribution, which is proportional to
$\exp(-\|\cdot\|_1)$.  A basic property of the scalar Laplace
distribution is that it is memory free. That is, given a Laplace
distribution with mean $1/\alpha$, then the probability relationship
\begin{equation}
\label{memoryFree}
\Pr(|r| > t_2 \mid |r| > t_1) = \Pr(|r| > t_2 - t_1) = \exp(-\alpha[t_2 - t_1])
\end{equation}
holds for all $t_2>t_1$. Hence, the probability that a scalar
residual is at least $2t$ away from the mean, given that it is at
least $t$ away from the mean, decays exponentially fast with $t$.
For large $t$, it is unintuitive to make such a strong claim for a
residual already known to correspond to an outlier.

Contrast this behavior with that of the Student's t-distribution.
When $\nu = 1$, the Student's t-distribution is simply the Cauchy
distribution, with a density proportional to $1/(1 + r^2)$.  Then we
have that
\[
 \lim_{t\to\infty} \Pr(|r|>2t \mid |r|>t) =
 \lim_{t\to\infty} \frac{\frac{\pi}{2}-\arctan(2t)}{\frac{\pi}{2} - \arctan(t)}
 = \frac12.
\]
Remarkably, the conditional probability is independent of $t$ for
large residuals. This cannot be achieved with any probability density
arising from a convex penalty, because~\eqref{memoryFree} provides a
lower bound for this family of densities, as is shown in the following
theorem.

\begin{btheorem}
\label{UniversalLowerBound}
Consider any scalar density $p$ arising from a symmetric proper closed convex
penalty $\rho$ via $p(t) =
\exp(-\rho(t))$, and take any point $t_0$ with positive right derivative
 $\alpha_0 = \partial_+\rho(t_0) > 0$.
Then for all $t_2 > t_1\geq t_0$, the conditional tail
distribution induced by $p(r)$ satisfies
\begin{equation*}
\label{memoryFreeIneq}
\Pr(|r| > t_2 \mid |r| > t_1) \leq \exp(-\alpha_0[t_2 - t_1])\;.
\end{equation*}
\end{btheorem}
\begin{proof}
  Define $\ell(t) = \rho(t_1) + \alpha_1(t - t_1)$ to be the (global)
  linear under-estimate for $\rho$ at $t_1$, where
  $\alpha_1= \partial_+\rho(t_1)$ is the right derivative of $\rho$ at
  $t_1$. Define $F(t) = \int_{t}^{\infty}p(r)\, dr$. We first note
  that $F(t)$ is log-concave (apply~\cite[Theorem 3]{Prekopa1971},
  taking the set $A = \{z \mid z \geq 0\}$).  Then $\log(F(t))$ is
  concave, and so its derivative
  \[
 \log(F(t))' =  \frac{p(t)}{-F(t)}
  \]
  is non-increasing. Therefore, the ratio $p(t)/F(t)$ is nondecreasing,
   and in particular
  \[
  \frac{p(t_1)}{F(t_1)} \leq \frac{p(t_2)}{F(t_2)}\;,
  \textt{or equivalently,}
  \frac{F(t_2)}{F(t_1)} \leq \frac{p(t_2)}{p(t_1)}\;.
  \]
  By assumption on the functions $\ell$ and $\rho$,
  \[
  \rho(t_2) - \ell(t_2) \geq \rho(t_1) - \ell(t_1) = 0,
  \]
  which implies that
  \[
  \begin{aligned}
    \Pr(|r| > t_2 \mid |r| > t_1)
    &= \frac{ F(t_2)}{F(t_1)} \leq
      \frac{\exp(-\rho(t_2))}{\exp(-\rho(t_1))}
   \\& = \exp(-[\rho(t_2) - \ell(t_1)])
   \\&\leq \exp(-[\ell(t_2) - \ell(t_1)])
   \\&= \exp(-\alpha_1[t_2 - t_1])\,.
\end{aligned}
\]
To complete the proof, note that the right derivative
$\partial_+\rho(t)$ is nondecreasing~\cite[Theorem 24.1]{RTR}. Then we
have $\alpha_0 \leq \alpha_1$ for $t_0 \leq t_1$.
\end{proof}

For differentiable log-concave densities, the
\emph{influence function} is defined to be $\rho'(t)$, and for a
general distribution it is the derivative of the negative log of the
density. These functions provide further insight into the difference
between the behaviors of log-concave densities and heavy-tailed
densities such as the Student's. In particular, they measure the
effect of the size of a residual on the negative log likelihood.
The Student's t-density has a so-called {\it redescending} influence
function: as residuals grow larger, they are effectively ignored by
the model.  Figure~\ref{fig:distributions} shows the relationships
among densities, penalties, and influence functions of two
log-concave distributions (Gaussian and Laplacian) and those of the
Student's t, which is not log-concave. If we examine the derivative
\begin{equation*}
  \rho'(r)=\frac{2r}{\nu+r^2}
\end{equation*}
of the Student's t-penalty \eqref{eq:19}, it is clear that large
residuals have a small influence when $r^2\gg \nu$. For small $r$, on
the other hand, the derivative resembles that of the least-squares
penalty.  See Hampel et al.~\cite{Hampel} for a discussion of
influence-function approaches to robust statistics, and redescending
influence functions in particular, and Shevlyakov et
al.~\cite{Shevlyakov2008} for further connections.

There is an implicit tradeoff between convex and non-convex penalties
(and their log-concave and non-log-concave counterparts). Convex
models are easier to characterize and solve, but may be wrong in a
situation in which large outliers are expected.
Nonconvex penalties are particularly useful with large outliers.

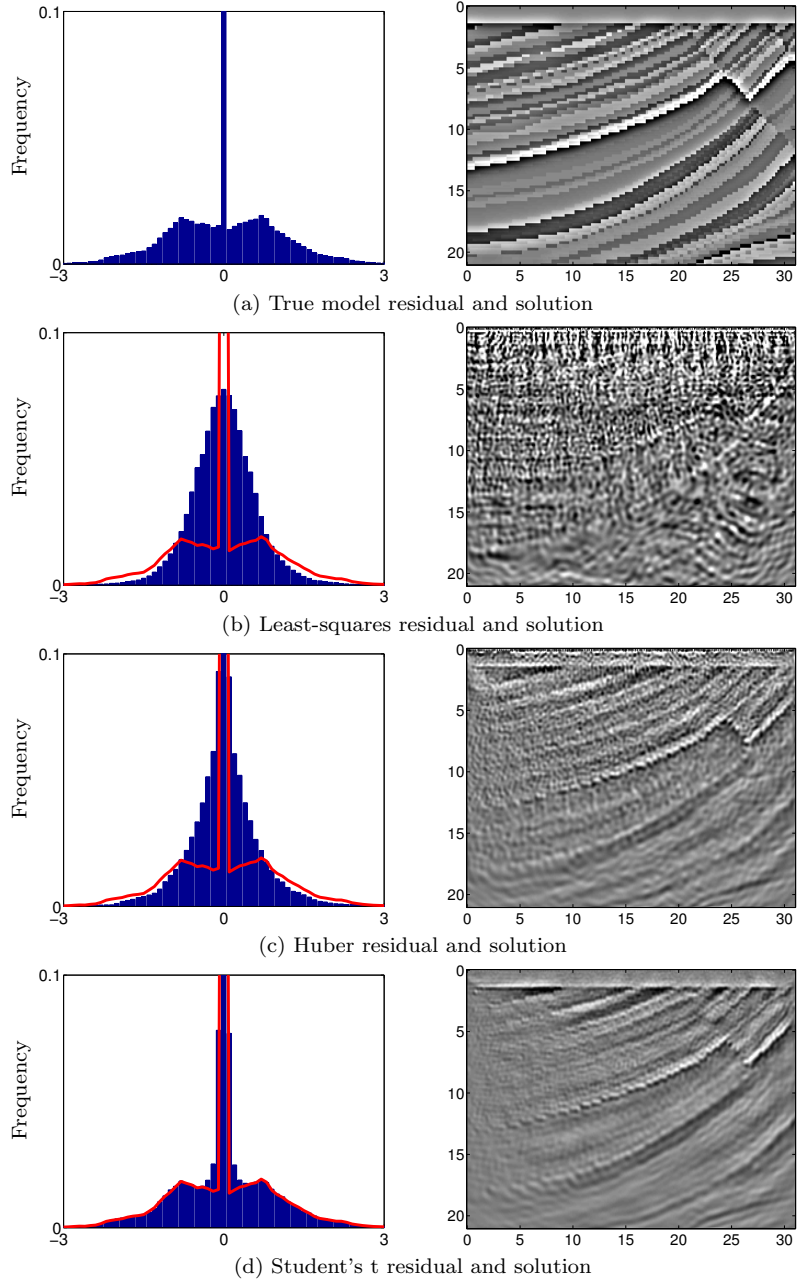
\begin{figure}
  \centering
  \subfloat[True model residual and solution]{%
    \label{fig:fwiresids_true}
    \includegraphics[width=.38\textwidth]{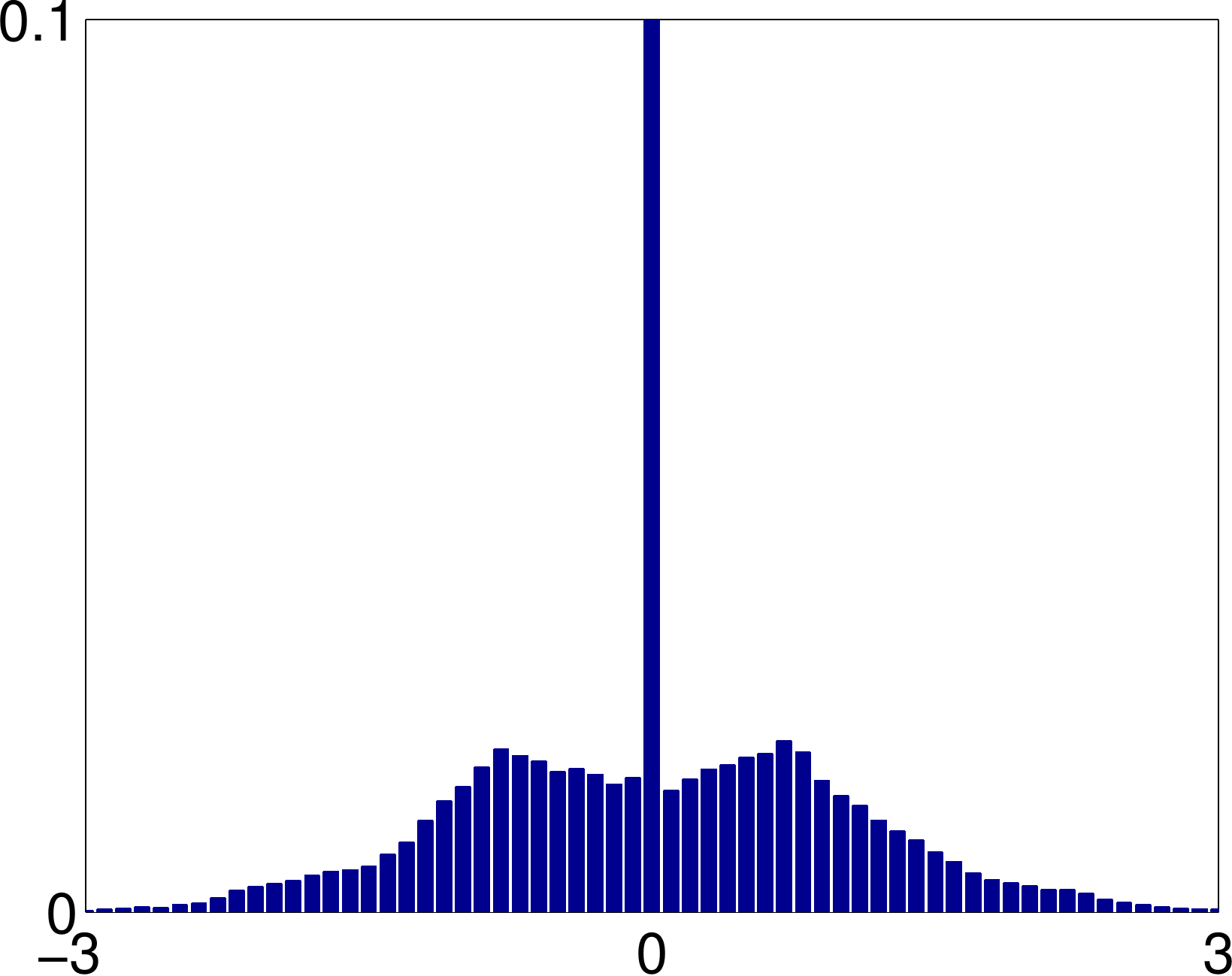}
    \begin{picture}(0,0)
      \put(-145,40){\rotatebox{90}{Frequency}}
    \end{picture}
    \qquad
    \includegraphics[width=.38\textwidth]{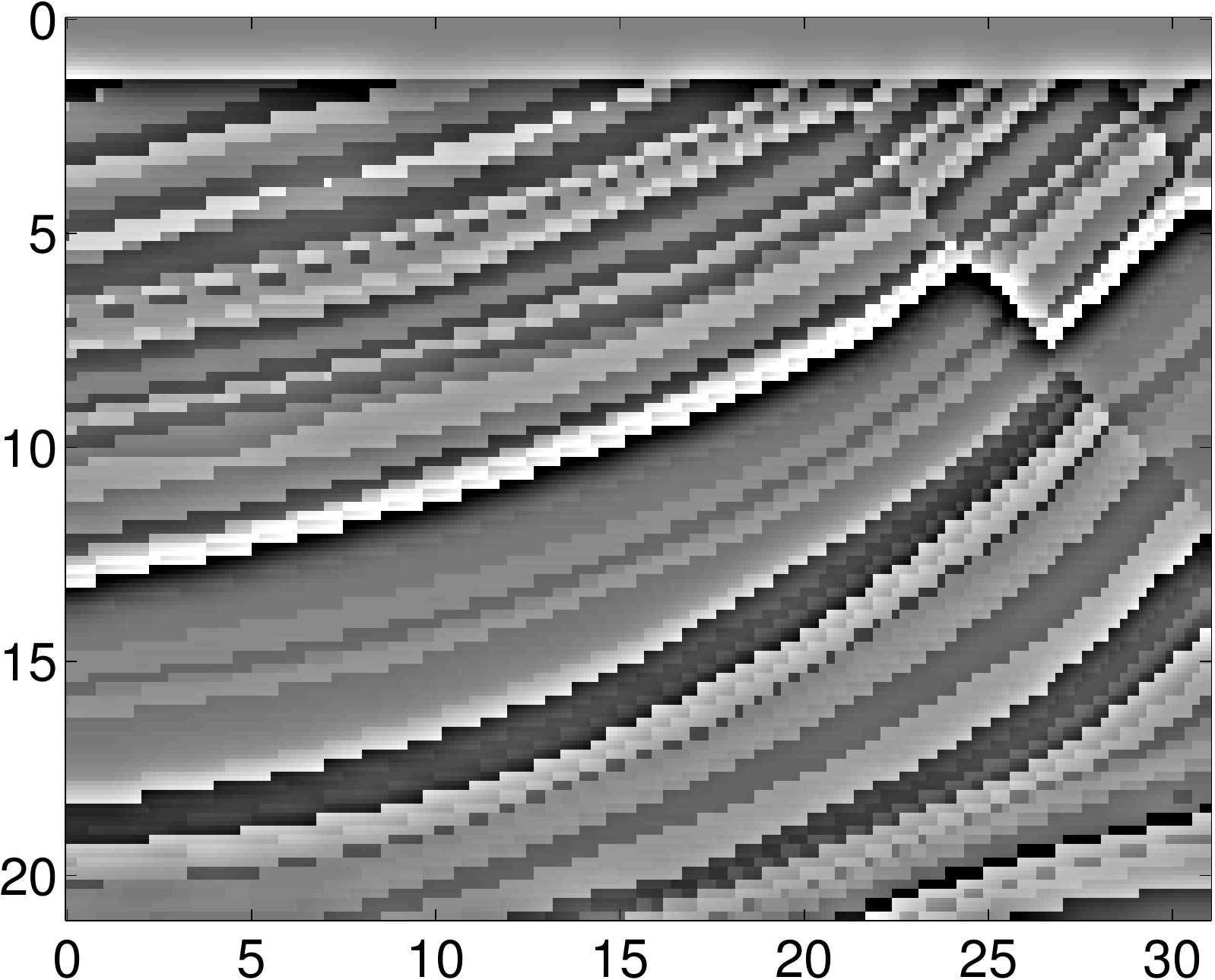}}
  \\[1pt]
  \subfloat[Least-squares residual and solution]{%
    \label{fig:fwiresids_ls}
    \includegraphics[width=.38\textwidth]{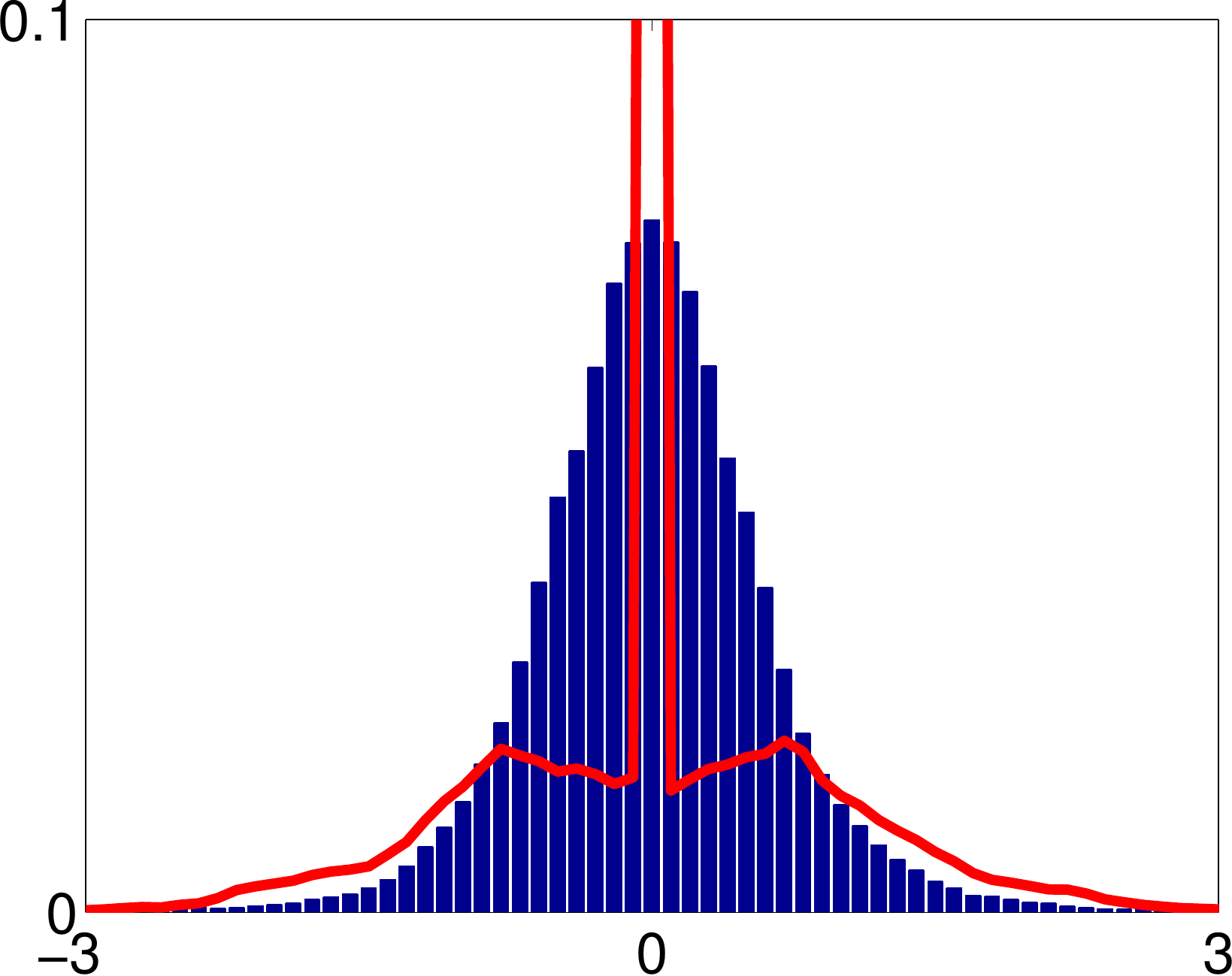}
    \begin{picture}(0,0)
      \put(-145,40){\rotatebox{90}{Frequency}}
    \end{picture}
    \qquad
    \includegraphics[width=.38\textwidth]{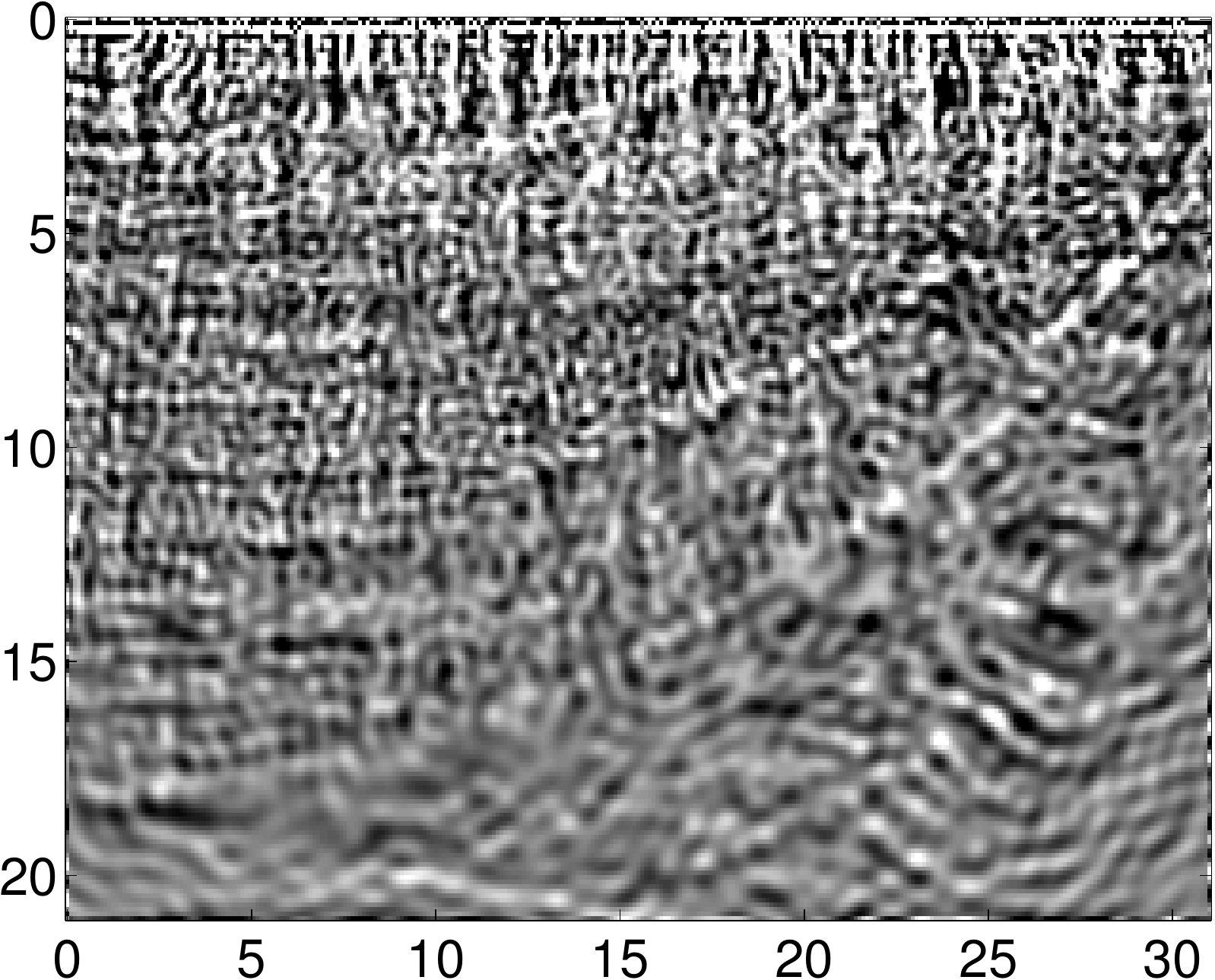}}
  \\[1pt]
  \subfloat[Huber residual and solution]{%
    \label{fig:fwiresids_huber}
    \includegraphics[width=.38\textwidth]{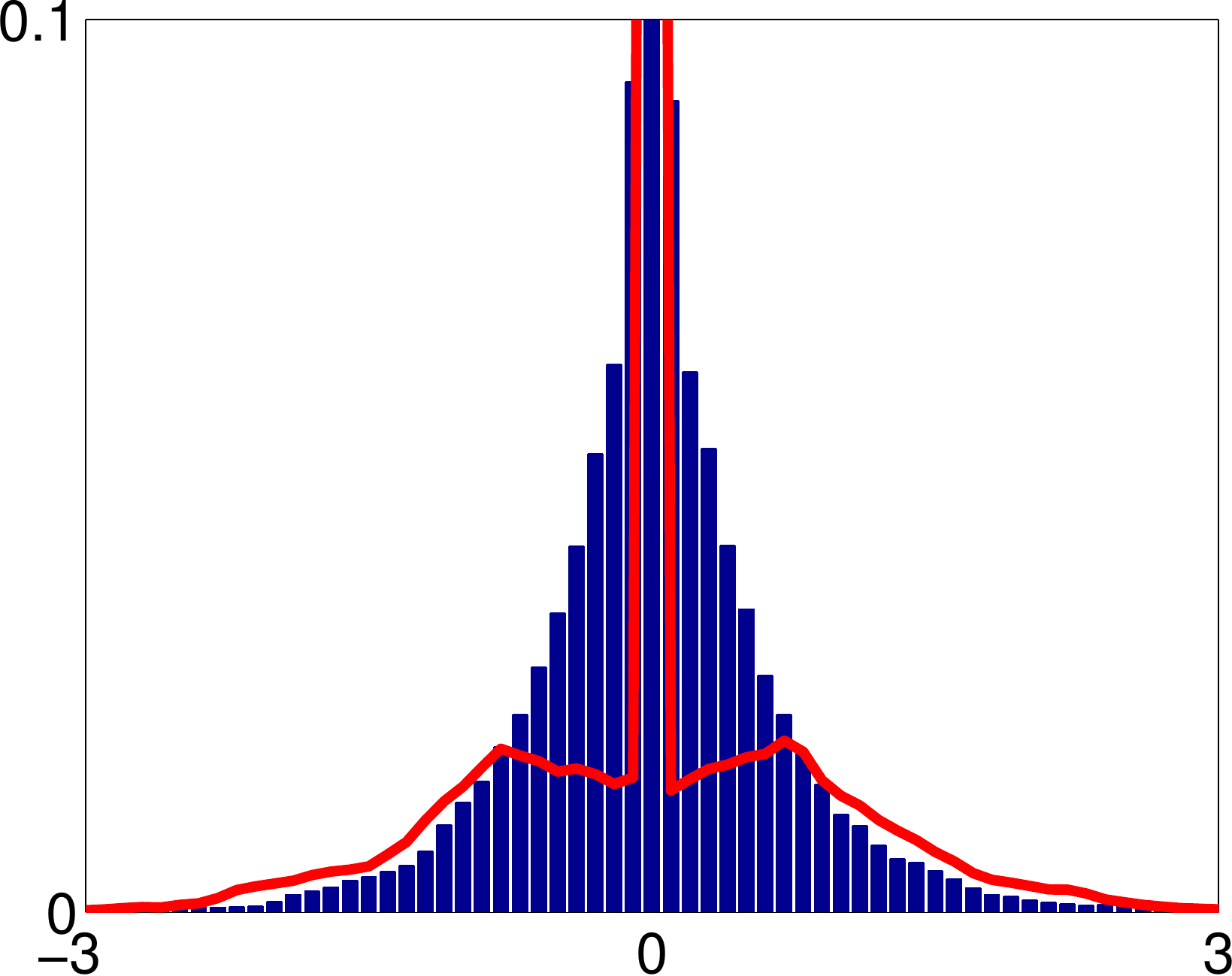}
    \begin{picture}(0,0)
      \put(-145,40){\rotatebox{90}{Frequency}}
    \end{picture}
    \qquad
    \includegraphics[width=.38\textwidth]{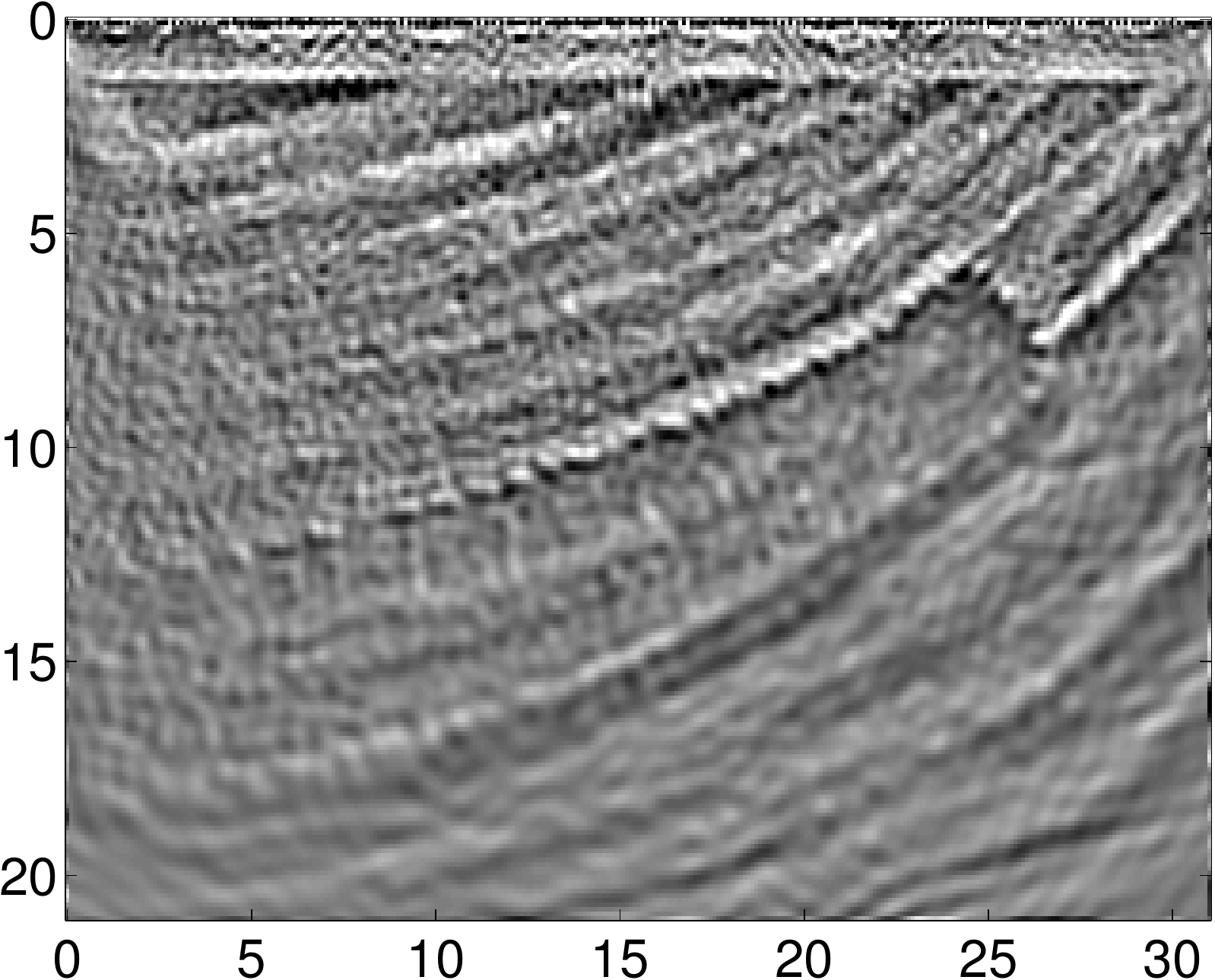}}
  \\[1pt]
  \subfloat[Student's t residual and solution]{%
    \label{fig:fwiresids_st}
    \includegraphics[width=.38\textwidth]{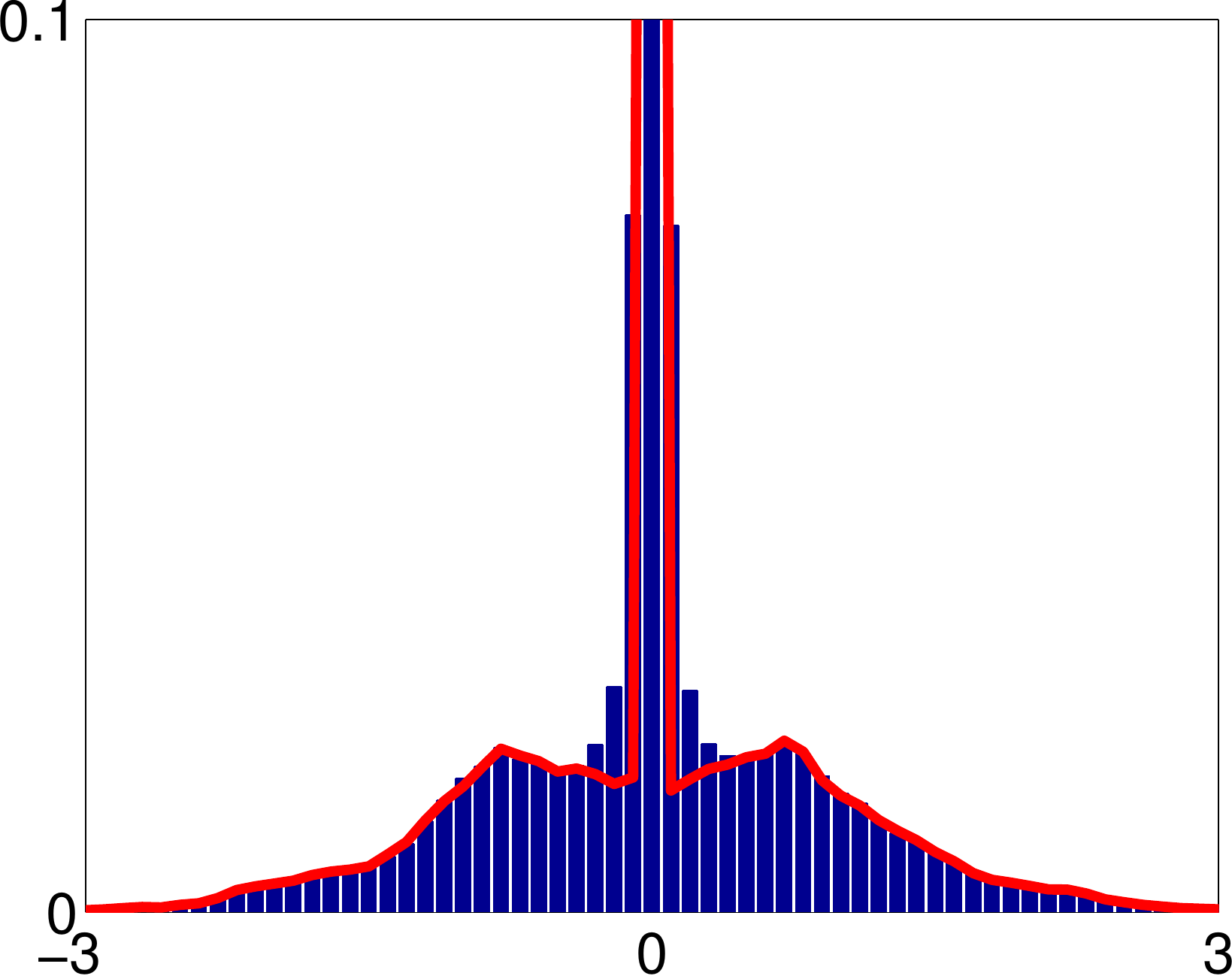}
    \begin{picture}(0,0)
      \put(-145,40){\rotatebox{90}{Frequency}}
    \end{picture}
    \qquad
    \includegraphics[width=.38\textwidth]{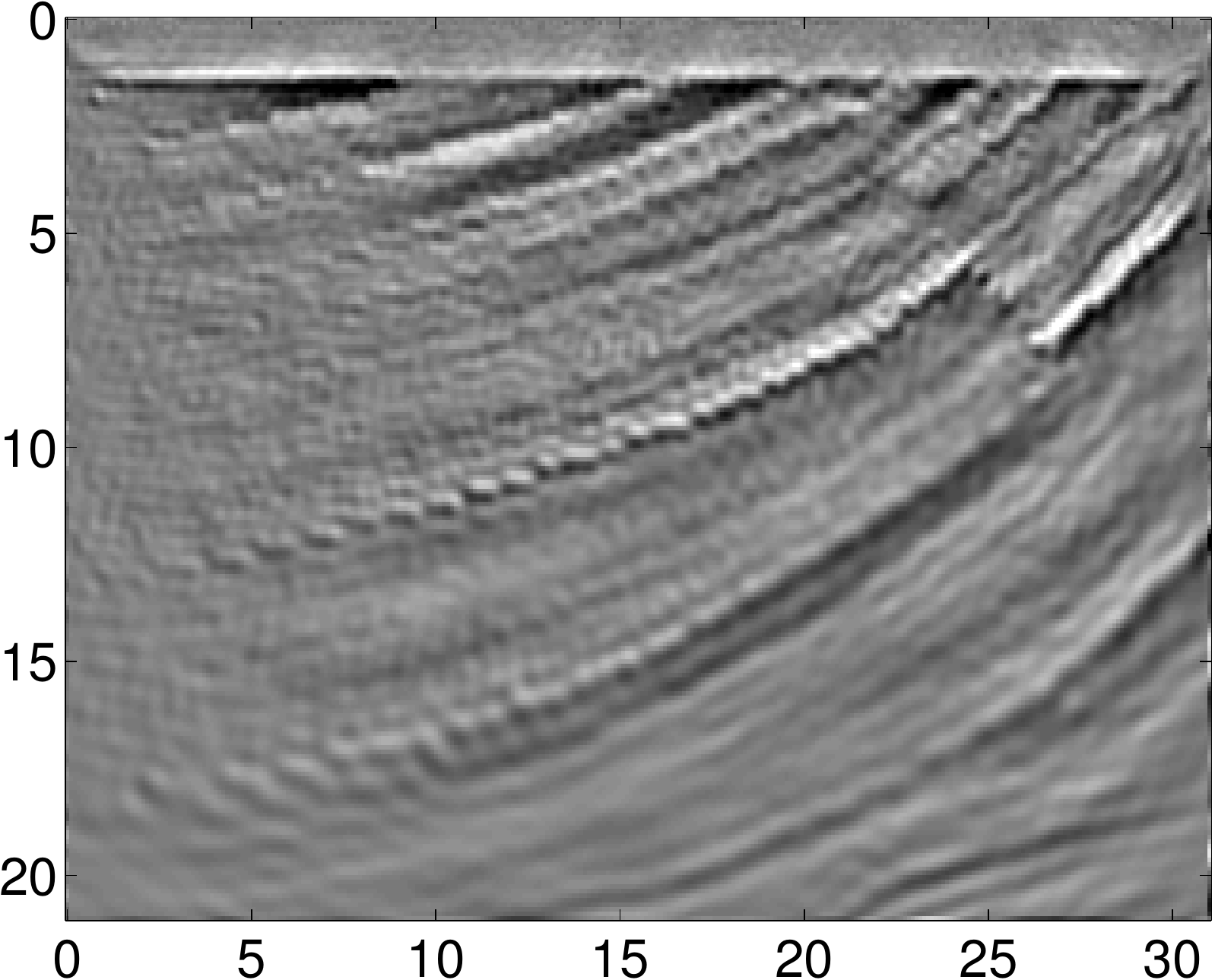}}
  \caption{Residual histograms (normalized) and solutions for an
    FWI problem.  The histogram at
    \protect\subref{fig:fwiresids_true} the true solution shows that the
    errors follow a tri-modal distribution (superimposed on the other
    histogram panels for reference).  The residuals for
    \protect\subref{fig:fwiresids_ls} least-squares and
    \protect\subref{fig:fwiresids_huber} Huber reconstructions follow
    the model error densities (i.e., Gaussian and Laplace). The
    residuals for \protect\subref{fig:fwiresids_st} the Student t
    reconstruction, however, closely match the distribution of the
    actual errors.}
   \label{fig:fwiresids}
\end{figure}

\subsection{The Student's t in practice}

Figure~\ref{fig:fwiresids} compares the reconstruction obtained using
the Student's t-penalty, with those obtained using least-squares and
Huber penalties, on an FWI experiment (described more fully in
\S\ref{sec:full-wavef-invers}). These panels show histograms of the
residuals~\eqref{eq:2} that are obtained at different solutions,
including the true solution, and the solutions recovered by
solving~\eqref{eq:5} where the subfunctions $\phi_i$ in~\eqref{eq:18}
are defined by the least-squares, Huber, and Student's t- penalties.

The experiment simulates 50\% missing data using a random mask that
zeros out half of the data obtained via a forward model at the true
value of $x$.  A residual histogram at the true $x$ therefore contains
a large spike at $0$, corresponding to the residuals for correct data,
and a multimodal distribution of residuals for the erased data.  The
least-squares recovery yields a residual histogram that resembles a
Gaussian distribution. The corresponding inversion result is useless,
which is not surprising, because the residuals at the true solution are
very far from Guassian.  The reconstruction using the Huber penalty is a
significant improvement over the conventional least-squares approach,
and the residual has a shape that resembles the Laplace distribution,
which is closer to the shape of the true residual. The Student's t
approach yields the best reconstruction, and, remarkably, produces a
residual distribution that matches the multi-modal shape of the true
residual histogram.  This is surprising because the Student's
t-distribution is unimodal, but the residual shape obtained using the
inversion formulation is not. It appears that the statistical prior
implied by the Student's t-distribution is weak enough to allow the
model to converge to a solution that is almost fully consistent with
the good data, and completely ignors the bad data.

Despite several successful applications in statistics and control
theory~\cite{Lange1989,Fahr1998}, Student's t-formulations do not
enjoy widespread use, especially in the context of nonlinear
regression and large-scale inverse problems.  Recently, however, they
were shown to work very well for robust recovery in nonlinear inverse
problems such as Kalman smoothing and bundle
adjustment~\cite{AravkinThesis2010}, and to outperform the Huber
penalty when inverting large synthetic
models~\cite{AravkinLH:2011a,AravkinLH:2011b}. Moreover, because the
corresponding penalty function is smooth, it is usually possible to
adapt existing algorithms and workflows to work with a robust
formulation.

In order for algorithms to be useful with industrial-scale problems,
it is essential that they be designed for conventional and robust
formulations that use a relatively small portion of the data in any
computational kernel.  We lay the groundwork for these algorithms in
the next section.

\section{Sample average approximations}
\label{sec:sampling}

The data-averaging approach used to derive the
approximation~\eqref{eq:3} may not be appropriate when the misfit
functions $\phi_i$ are something other than the 2-norm. In particular,
a result such as Proposition~\ref{prop:generalSampling}, which
reassures us that the approximations are unbiased estimates of the
true functions, relies on the special structure of the 2-norm, and is
not available to us in the more general case. In this section, we
describe sampling strategies---analogous to the stochastic-trace
estimation procedure of \S\ref{sec:dr}---that allow for more general
misfit measures $\phi_i$. In particular, we are interested in a
sampling approach that allows for differential treatment across
experiments $i$, and for robust functions.

We adopt the useful perspective that each of the constituent functions
$\phi_i$ and the gradients $\nabla\phi_i$ are members of a fixed
population of size $m$. The aggregate objective function and its
gradient,
\[
 \phi(x) = \frac1m\sum_{i=1}^m\phi_i(x)
 \text{and}
 \nabla\phi(x) = \frac1m\sum_{i=1}^m \nabla\phi_i(x),
\]
can then simply be considered to be population averages of the
individual objectives and gradients, as reflected in the scaling factors
$1/m$.  A common method for estimating the mean of a population is
to sample only a small subset $\sample\subseteq\Set{1,\ldots,m}$ to
derive the sample averages
\begin{equation}\label{eq:sample-phi}
 \phis(x) = \frac1{s}\sum_{i\in\sample}\phi_i(x)
 \text{and}
 \nabla\phis(x) = \frac1{s}\sum_{i\in\sample}\nabla\phi_i(x),
\end{equation}
where $s=|\sample|$ is the sample size.  We build the subset $\sample$
as a uniform random sampling of the full population, and in that case
the sample averages are unbiased:
\begin{equation}\label{eq:g-sample-avg}
\expval[\phis(x)] = \phi(x) \text{and} \expval[\nabla\phis(x)] = \nabla\phi(x).
\end{equation}

The cost of evaluating these sample-average approximations is about
$s/m$ times that for the true function and gradient.  (Non-uniform
schemes, such as importance and stratified sampling, are also
possible, but require prior knowledge about the relative importance of
the $\phi_i$.)  We use these quantities to drive the optimization
procedure.

This approach constitutes a kind of dimensionality-reduction scheme,
and it is widely used by census takers to avoid the expense of
measuring the entire population. In our case, measuring each element of
the population means an evaluation of a function~$\phi_i$ and its
gradient $\nabla\phi_i$. The goal of probability sampling is to design
randomized sampling schemes that estimate statistics---such as these
sample averages---with quantifiable error; see, for example, Lohr's
introductory text~\cite{lohr1999sampling}.

The stochastic-optimization methods that we describe
in~\S\ref{sec:semistochastic} allow for approximate gradients, and thus
can take advantage of these sampling schemes. The error analysis of
the sample-average method described in \S\ref{sec:sampling-opt} relies
on the second moment of the error
\begin{equation} \label{eq:sample-e}
 e = \nabla\phis - \nabla\phi
\end{equation}
in the gradient.  Because the sample averages are unbiased, the
expected value of the squared error of the approximation reduces to
the variance of the norm of the sample average:
\begin{equation} \label{eq:var-sample-avg}
\expval\big[\norm{e}^2\big] = \var\big[ \norm{\nabla\phis} \big].
\end{equation}
This error is key to the optimization process, because the accuracy of
the gradient estimate ultimately determines the quality
of the search directions available to the underlying optimization
algorithm.

\subsection{Sampling with and without replacement}

Intuitively, the size $s$ of the random sample influences the norm of
the error $e$ in the gradient estimate. The difference between uniform
sampling schemes with or without replacement greatly affects how the
variance of the sample average decreases as the sample size
increases. In both cases, the variance of the estimator is
proportional to the sample variance
\begin{equation} \label{FiniteGradientSampleVariance}
 \sigma_g^2 := \frac{1}{m-1}\sum_{i=1}^m \norm{\nabla\phi_i - \nabla\phi}^2
\end{equation}
of the population of gradients $\Set{\nabla\phi_1,\ldots,
  \nabla\phi_m}$ evaluated at $x$.  This quantity is inherent to the
problem and independent of the chosen sampling scheme.

When sampling from a finite population without replacement (i.e.,
every element in $\sample$ occurs only once), then the error $e_n$ of
the sample average gradient satisfies
\begin{equation} \label{WithoutReplacementVariance}
  \expval[\norm{e_n}^2] = \frac1s\left(1 - \frac{s}{m}\right) \sigma^2_g\,;
\end{equation}
for example, see Cochran~\cite{Cochran1977} or
Lohr~\cite[\S2.7]{lohr1999sampling}.  Note that the expected error
decreases with $s$, and---importantly---is exactly 0 when $s=m$.  On
the other hand, in a sample average gradient built by uniform sampling with
replacement, every sample draw of the population is independent of the
others, so that the error $e_r$ of this sample average gradient satisfies
\begin{equation}
  \label{WithReplacementVariance}
  \expval[\norm{e_r}^2] = \frac{1}{s}\sigma^2_g.
\end{equation}
This error goes to 0 as $1/s$, and is never 0 when sampling over a
finite population.

Comparing the expected error between sampling with and without
replacement for finite populations, we note that
\[
\expval[\norm{e_n}^2] = \left(1-\frac{s}{m}\right)\expval[\norm{e_r}^2],
\]
and so sampling without replacement yields a uniformly lower expected
error than independent finite sampling.

\subsection{Data averaging}

The data-averaging approach discussed in~\S\ref{sec:dr} for the
objective~\eqref{eq:17} does not immediately fit into the
sample-average framework just presented, even though the function
$\phi\W$ defined in~\eqref{eq:3} is a sample average. Nevertheless,
for all sampling schemes described by
Proposition~\ref{prop:generalSampling}, the sample average
\[
\phi\W(x) = \frac1s\sum_{j=1}^s\phitilde_i(x),
\text{with}
\phitilde_i(x) := \|R(x)w_i\|^2,
\]
is in some sense a sample average of an infinite population. If the
random vectors are uncorrelated---as required by
Proposition~\ref{prop:generalSampling}---than, as
with~\eqref{WithReplacementVariance}, the error \[e_w = \nabla\phi\W -
\phi\] of the sample average gradient is proportional to the sample
variance of the population of gradients of $\phi\W$. That is,
\[
\expval[\norm{e_w}^2] = \frac1s\sigmatilde_g^2,
\]
where $\sigmatilde_g^2$ is the sample variance of the population of
gradients $\{\,\nabla\phitilde_1,\ldots,\nabla\phitilde_m\,\}$.

The particular value of $\sigmatilde_g^2$ will depend on the distribution
from which the weights $w_i$ are drawn; for some distributions of
$w_i$ this quantity may even be infinite, as is shown by the
following results.

The sample variance~\eqref{FiniteGradientSampleVariance} is always
finite, and the analogous sample variance $\sigmatilde_g^2$ of the
implicit functions $\nabla\phitilde_i$ is finite under general
conditions on $w$.

\begin{bproposition}
  \label{FiniteVarianceProposition}
  The sample variance $\sigmatilde_g^2$ of the population
  $\{\,\nabla\phitilde_1,\ldots,\nabla\phitilde_m\,\}$ of gradients is
  finite when the distribution for $w_i$ has finite fourth moments.
\end{bproposition}
\begin{proof}
  The claim follows from a few simple bounds (all sums run from 1 to
  $m$):
 \[
 \begin{aligned}
   \sigmatilde_g^2
   &\leq \expval\left[\|\nabla\phitilde_{i}\|^2\right]
 \\&=
   4\expval
   \left[\,
     \left\|
       \left( \sum_i w_i \nabla  r_i(x)  \right)
       \left( \sum_i w_i r_i(x)         \right)
     \right\|^2
   \right]
 \\&\leq
  4 \expval
   \left[\,
     \left(\left\|
        \sum_i w_i \nabla  r_i(x) \right\|_2
\left\|       \sum_i w_i r_i(x)
     \right\|\right)^2
   \right]
\\&\leq
  4 \expval
   \left[\,
     \left(\sum_i\left\|
        w_i \nabla  r_i(x)  \right\|_2
\sum_i\left\|      w_i r_i(x)
     \right\|\right)^2
   \right]
\\ &=
  4 \expval
   \left[\,
     \left(\sum_i|w_i|\left\|
         \nabla  r_i(x)  \right\|_2
\sum_i |w_i|\left\|      r_i(x)
     \right\|\right)^2
   \right]
 \\&\leq
 4\max_i m^2\|\nabla r_i(x)\|_2^2\cdot\max_i\|r_i(x)\|^2
  \expval \bigg[\sum_{ij}w_i^2w_j^2\bigg]\;.
 \end{aligned}
 \]
 The quantity $\expval \big[\sum_{ij}w_i^2w_j^2\big] <
 \infty$ when the fourth moments are finite.
\end{proof}

As long as $\sigmatilde_g^2$ is nonzero, the expected error of uniform
sampling without replacement is asymptotically better than the
expected error that results from data averaging. That is,
\[
  \expval[\norm{e_n}^2] < \expval[\norm{e_w}^2]
  \quad
  \hbox{for all $s$ large enough.}
\]
At least as measured by the second moment of the
error in the gradient, the simple random sampling without replacement
has the benefit of yielding a good estimate when compared to these
other sampling schemes.

\section{Stochastic optimization}
\label{sec:semistochastic}

Stochastic optimization, which naturally allows for inexact
gradient calculations, meshes well with the various sampling
and averaging strategies described in \S\ref{sec:sampling}. We review
several approaches that fall under the stochastic optimization
umbrella, and describe their relative benefits.

Although the full-waveform inversion application that we consider is
nonconvex, the following discussion make the assumption that the
optimization problem is convex; this expedient concedes the analytical
tools that allow us to connect sampling with rates of convergence. It
is otherwise difficult to connect a convergence rate to the sample
size; see~\cite[\S2.3]{FS:2011}.  The usefulness of the approach is
justified by numerical experiments, both in the present paper and
in~\cite[\S5]{FS:2011}, where results for both convex and nonconvex
models are presented.

\subsection{Stochastic gradient methods}
\label{sec:stochastic}

Stochastic gradient methods for minimizing a differentiable function
$\phi$, not necessarily of the form defined in~\eqref{eq:5}, can be
generically expressed by the iteration
\begin{equation}
  \label{eq:sgIter}
  x\kp1 = x\k - \alpha\k d\k \text{with} d\k := s\k + e\k,
\end{equation}
where $\alpha\k$ is a positive stepsize, $s\k$ is a descent direction
for $\phi$, and $e\k$ is a random noise term.  Bertsekas and
Tsitsiklis~\cite[Prop.~3]{BT:2000} give general conditions under which
\[
\lim_{k\to\infty}\nabla \phi(x\k)=0,
\]
and every limit point of $\{x_k\}$ is a stationary point of $\phi$.
Note that unless the minimizer is unique, this does not imply that the
sequence of iterates $\{x\k\}$ converges. Chief among the required
conditions are that $\nabla \phi$ is globally Lipschitz, i.e., for
some positive $L$,
\begin{equation*}
 \norm{\nabla\phi(x)-\nabla\phi(y)}\le L\norm{x-y}
  \text{for all $x$ and $y$;}
\end{equation*}
that for all $k$,
\begin{subequations} \label{eq:stoch-reqs}
\begin{align}
  \label{eq:grad-related-1}
 s\k\T\nabla \phi(x\k) \le -\mu_1\norm{\nabla \phi(x\k)}^2,
 \\ \label{eq:grad-related-2}
 \norm{s\k} \le \mu_2(1+\norm{\nabla \phi(x\k)}),
 \\\label{eq:NoiseBound}
 \expval[e\k]=0 \text{and} \expval\big[\norm{e\k}^2\big] < \mu_3,
\end{align}
\end{subequations}
for some positive constants $\mu_1$, $\mu_2,$ and $\mu_3$; and that
the steplengths satisfy the infinite travel and summable conditions
\begin{equation}\label{eq:6}
  \sum_{k=0}^\infty \alpha\k=\infty
  \text{and}
  \sum_{k=0}^\infty \alpha_k^2<\infty.
\end{equation}
Many authors have worked on similar stochastic-gradient methods, but
the Bertsekas and Tsitsiklis~\cite{BT:2000} is particularly general;
see their paper for further references.

Note that the randomized sample average schemes (with or without
replacement) from \S\ref{sec:sampling} can be immediately used to
design a stochastic gradient that
satisfies~\eqref{eq:grad-related-2}. It suffices to choose the sample
average of the gradient~\eqref{eq:sample-phi} as the search direction:
\[
d\k = \nabla{\phis}(x\k).
\]
Because the sample average $\nabla\phis$ is
unbiased---cf. \eqref{eq:g-sample-avg}---this direction is on average
simply the steepest descent, and can be interpreted as having been
generated from the choices
\[
s\k = \nabla\phi(x\k) \text{and} e\k = \nabla\phis(x\k) - \nabla\phi(x\k).
\]
Moreover, the sample average has finite
variance---cf.~\eqref{WithoutReplacementVariance}--\eqref{WithReplacementVariance}---and
so the direction $s\k$ and the error $e\k$ clearly satisfy
conditions~\eqref{eq:stoch-reqs}.

The same argument holds for the data-averaging scheme outlined in
\S\ref{sec:dr}, as long as the distribution of the mixing vector
admits an unbiased sample average with a finite
variance. Propositions~\ref{prop:generalSampling}
and~\ref{FiniteVarianceProposition} establish conditions under which
these requirements hold.

Suppose that $\phi$ is strongly
convex with parameter $\mu$, which implies that
\begin{equation*}
 \frac{\mu}2\norm{x\k-\xstar}^2 \le \phi(x\k) - \phi(\xstar),
\end{equation*}
where $\xstar$ is the unique minimizer of $\phi$. Under this
additional assumption, further statements can be made about the rate
of convergence. In particular, the iteration \eqref{eq:sgIter}, with
$s\k=\nabla\phi(x\k)$, converges sublinearly, i.e.,
\begin{equation}\label{eq:9}
  \expval[\norm{x\k - \xstar}] = \Oscr(1/k).
\end{equation}
where the steplengths $\alpha\k=\Oscr(1/k)$ are decreasing
\cite[\S2.1]{nemirovski2009robust}. This is in fact the optimal rate
among all first-order stochastic
methods~\cite[\S14.1]{nemirovski1994efficient}.

A strength of the stochastic algorithm~\eqref{eq:sgIter} is that it
applies so generally. All of the sampling approaches that we have
discussed so far, and no doubt others, easily fit into this
framework. The convergence guarantees are relatively weak for our
purposes, however, because they do not provide guidance on how a
sampling strategy might influence the speed of convergence. This
analysis is crucial within the context of the sampling schemes that we
consider, because we want to gain an understanding of how the sample
size influences the speed of the algorithm.

\subsection{Incremental-gradient methods}
\label{sec:incremental-grad}

Incremental-gradient methods, in their randomized form, can be
considered a special case of stochastic gradient methods that are
especially suited to optimizing sums of functions such
as~\eqref{eq:5}. They can be described by the iteration scheme
\begin{equation}
  \label{eq:4}
  x\kp1 = x\k - \alpha\k\nabla\phi_{i_k}(x\k),
\end{equation}
for some positive steplengths $\alpha\k$, where the index $i_k$
selects among the $m$ constituent functions of $\phi$. In the
deterministic version of the algorithm, the ordering of the
subfunctions $\phi_i$ is predetermined, and the counter $i_k = (k\bmod
m)+1$ makes a full sweep through all the functions every $m$
iterations. In the randomized version, $i_k$ is at each iteration
randomly selected with equal probability from the indices
$1,\ldots,m$.  (The Kaczmarz method for linear
system~\cite{kaczmarz1937aav} is closely related, and a randomized
version of it is analyzed by Strohmer and
Vershynin~\cite{strohmer2009randomized}.)

In the context of the sampling discussion in \S\ref{sec:sampling}, the
incremental-gradient algorithm can be viewed as an extreme sampling
strategy that at each iteration uses only a single function $\phi_i$
(i.e., a sample of size $s=1$) in order to form a sample average
$\phis$ of the gradient. For the data-averaging case of
\S\ref{sec:dr}, this corresponds to generating the approximation
$\phi\W$ from a single weighted average of the data (i.e., using a
single random vector $w_i$ to form $R(x)w_i$).

Bertsekas and Tsitsiklis~\cite[Prop.~3.8]{bertsekas1996neuro} describe
conditions for convergence of the incremental-gradient algorithm for
functions with globally Lipschitz continuous gradients, when the
steplengths $\alpha\k\to0$ as specified by~\eqref{eq:6}. Note that it
is necessary for the steplengths $\alpha\k\to0$ in order for the
iterates $x\k$ produced by~\eqref{eq:4} to ensure stationarity of the
limit points. Unless we assume that $\nabla\phi(\xbar)=0$ implies that
$\nabla\phi_i(\xbar)=0$ for all $i$, a stationary point of $\phi$ is not a
fixed point of the iteration process;
Solodov~\cite{solodov1998incremental} and Tseng~\cite{tseng:1998}
study this case.  Solodov~\cite{solodov1998incremental} further
describes how bounding the steplengths away from zero yields limit
points $\xbar$ that satisfy the approximate stationarity condition
\[
\norm{\nabla\phi(\xbar)}=\Oscr\big(\inf_k\alpha\k\big).
\]

With the additional assumption of strong convexity of $\phi$, it
follows from Nedi\'c and Bertsekas~\cite{nedic2000convergence} that
the randomized incremental-gradient algorithm with a decreasing
stepsize $\alpha\k=\Oscr(1/k)$ converges sublinearly accordingly
to~\eqref{eq:9}. They also show that keeping the stepsize constant as
$\alpha\k\equiv m/L$ implies that
\begin{equation*}
 \expval[\norm{x\k-\xstar}^2] \le \Oscr([1-\mu/L]^k) + \Oscr(m/L).
\end{equation*}
This expression is interesting because the first term on the
right-hand side decreases at a linear rate, and depends on the
condition number $\mu/L$ of $\phi$; this term is present for any
deterministic first-order method with constant stepsize. Thus, we can
see that with the strong-convexity assumption and a constant stepsize,
the incremental-gradient algorithm has the same convergence
characteristics as steepest descent, but with an additional constant
error term.

\subsection{Sampling methods}
\label{sec:sampling-opt}

The incremental-gradient method described in
\S\ref{sec:incremental-grad} has the benefit that each iteration costs
essentially the same as evaluating only a single gradient element
$\nabla\phi_i$. The downside is that they achieve only a sublinear
convergence to the exact solution, or a linear convergence to an
approximate solution. The sampling approach described in Friedlander
and Schmidt~\cite{FS:2011} allows us to interpolate between the
one-at-a-time incremental-gradient method at one extreme, and a full
gradient method at the other.

The sampling method is based on the iteration update
\begin{equation}
  \label{eq:10}
  x\kp1 = x\k - \alpha g\k, \quad \alpha = 1/L,
\end{equation}
where $L$ is the Lipschitz constant for the gradient, and the search
direction
\begin{equation}
  \label{eq:11}
  g\k = \nabla\phi(x\k) + e\k
\end{equation}
is an approximation of the gradient; the term $e\k$ absorbes the
discrepancy between the approximation and the true gradient.  We
define the direction $g\k$ in terms of the sample average
gradient~\eqref{eq:sample-phi}, and then $e\k$ corresponds to the
error defined in~\eqref{eq:sample-e}.

When the function $\phi$ is strongly convex and has a globally
Lipschitz continuous gradient, than the following theorem links the
convergence of the iterates to the error in the gradient.
\begin{btheorem} \label{th:g-w-error}
  Suppose that $\expval[\norm{e\k}^2] \le B\k$, where $\lim_{k\to\infty}
  B\kp1/B\k\le 1$. Then each iteration of algorithm~\eqref{eq:10}
  satisfies for each $k=0,1,2,\ldots,$
  \begin{equation}
    \label{eq:12}
    \expval[\norm{x\k-\xstar}^2] \le \Oscr([1-\mu/L]^k) + \Oscr(C\k),
  \end{equation}
  where $C\k = \max\{B\k,(1-\mu/L+\epsilon)^k\}$ for any positive $\epsilon$.
\end{btheorem}

It is also possible to replace $g\k$ in \eqref{eq:10} with a search
direction $p\k$ that is the solution of the system
\begin{equation}
  \label{eq:16}
  H\k p=g\k,
\end{equation}
for any sequence of Hessian approximations $H\k$ that are uniformly
positive definite and bounded in norm, as can be enforced in practice.
Theorem~\ref{th:g-w-error} continues to hold in this case, but with
different constants $\mu$ and $L$ that reflect the conditioning of the
``preconditioned'' function; see~\cite[\S1.2]{FS:2011}.

It is useful to compare~\eqref{eq:9} and~\eqref{eq:12}, which are
remarkably similar. The distance to the solution, for both the
incremental-gradient method \eqref{eq:4} and the gradient-with-errors
method \eqref{eq:10}, is bounded by the same linearly convergent
term. The second terms in their bounds, however, are crucially
different: the accuracy of the incremental-gradient method is bounded
by a multiple of the fixed steplength; the accuracy of the
gradient-with-errors method is bounded by the norm of the error in the
gradient.

Theorem~\ref{th:g-w-error} is significant because it furnishes a guide
for refining the sample $\sample\k$ that defines the average
approximation
\[
 g\k = \frac1{s\k}\sum_{i\in\sample\k}\phi_i(x\k)
\]
of the gradient of $\phi$, where $s\k$ is the size of the sample
$\sample\k$; cf. \eqref{eq:sample-phi}. In
particular,~\eqref{WithoutReplacementVariance}
and~\eqref{WithReplacementVariance} give the second moment of the
errors of these sample averages, which correspond precisely to the
gradient error defined by~\eqref{eq:11}.  If we wish to design a
sampling strategy that gives a linear decrease with a certain rate,
then a policy for the sample size $s\k$ needs to ensure that it grows
fast enough to induce $\expval[\norm{e\k}^2]$ to decrease with at
least that rate. Also, from~\eqref{eq:12}, it is clear that there is
no benefit in increasing the sample size at a rate faster than the
underlying ``pure'' first-order method without gradient error. If, for
example, the function is poorly conditioned---i.e., $\mu/L$ is
small---than the sample-size increase should be commensurately slow.

\begin{figure}[t]
  \centering
  \subfloat[]{%
    \label{fig:sampling-a}
    \includegraphics[trim=0 5 0 0,clip,width=.49\textwidth]{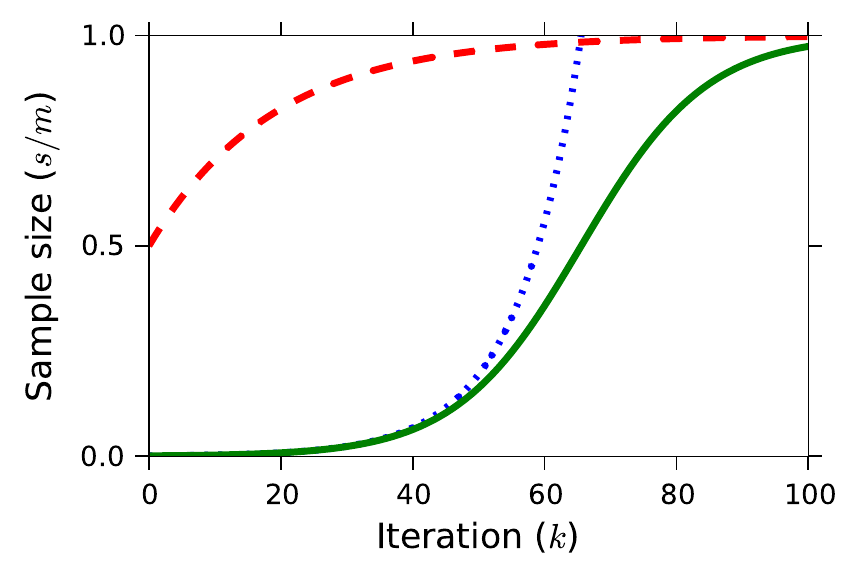}}
  \subfloat[]{%
    \label{fig:sampling-b}
    \includegraphics[trim=0 5 0 0,clip,width=.49\textwidth]{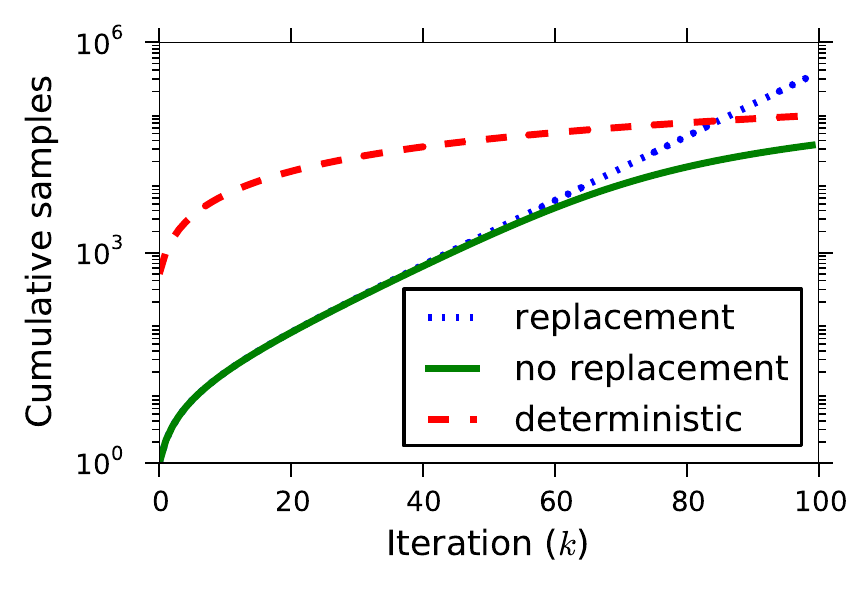}}
  \caption{Comparing the difference between the theoretical errors bounds in the sample
    averages for three sampling strategies (randomized with replacement, randomized without
    replacement, and deterministic). (a) Sample sizes, as fractions of the total population $m=1000$,
     required to reduce the error linearly with error constant 0.9. (b) The corresponding cumulative
    number of samples used. See bounds~\eqref{eq:14}.}
  \label{fig:sampling}
\end{figure}
It is instructive to compare how the sample average error decreases in the
randomized (with and without replacement) and deterministic cases. We
can more easily compare the randomized and deterministic variants by
following Bertsekas and Tsitsiklis~\cite[\S4.2]{bertsekas1996neuro},
and assuming that
\[
 \norm{\nabla\phi_i(x)}^2 \le \beta_1 + \beta_2\norm{\nabla\phi(x)}^2
 \quad
 \hbox{for all $x$ and $i=1,\ldots,m,$}
\]
for some constants $\beta_1\ge0$ and $\beta_2\ge1$. Together with the
Lipschitz continuity of $\phi$, we can provide the following bounds:
\begin{subequations}
  \label{eq:14}
  \begin{align}
   &\hbox{randomized, without replacement}&
    \expval[\norm{e\k}^2] &\le
    \frac{1}{s\k}\bigg[1-\frac{s\k}{m}\bigg]\bigg[\frac{m}{m-1}\bigg]\beta_k
    \\
    &\hbox{randomized, with replacement}&
    \expval[\norm{e\k}^2] &\le
    \frac{1}{s\k}\bigg[\frac{m}{m-1}\bigg]\beta_k
    \\
    &\hbox{deterministic}&
    \norm{e\k}^2\phantom] &\le
    4\bigg[\frac{m-s\k}{m}\bigg]^2\beta\k,
  \end{align}
\end{subequations}
where $\beta\k = \beta_1 + 2\beta_2 L [\phi(x\k) -
\phi(\xstar)]$. These bounds follow readily from the derivation in
\cite[\S\S3.1--3.2]{FS:2011}. Figure~\ref{fig:sampling} illustrates
the difference between these bounds on an example problem with
$m=1000$. The panel on the left shows how the sample size needs to be
increased in order for the right-hand-side bounds in~\eqref{eq:14} to
decrease linearly at a rate of 0.9. The panel on the right shows the
cumulative sample size, i.e., $\sum_{i=0}^{k} s_i$. Uniform sampling
without replacement yields a uniformly and significantly better bound
than the other sampling strategies.  Both types of sampling are
admissible, but sampling without replacement requires a much slower
rate of growth of $s$ to guarantee a linear rate.

The strong convexity assumption needed to derive the error bounds
used in this section is especially strong because the inverse
problem we use to motivate the sampling approach is not a convex
problem. In fact, it is virtually impossible to guarantee convexity of
a composite function such as~\eqref{eq:18} unless the penalty
function $\rho(\cdot)$ is convex and each $r_i(\cdot)$ is
affine. This is not the case for many interesting inverse problems,
such as full waveform inversion, and for nonconvex loss functions
corresponding to distributions with heavy tails, such as Student's t.

Even relaxing the assumption on $\phi$ from strong convexity to just
convexity makes it difficult to design a sampling strategy with a
certain convergence rate. The full-gradient method for convex (but not
strongly) functions has a sublinear convergence rate of
$\Oscr(1/k)$. Thus, all that is possible for a sampling-type approach
that introduces errors into the gradient is to simply maintain that
sublinear rate. For example, if $\norm{e\k}^2 \le B\k$, and
$\sum_{k=1}^\infty B\k<\infty$, then the iteration~\eqref{eq:10}
maintains the sublinear rate of the gradient
method~\cite[Theorem~2.6]{FS:2011}.  The theory for the strongly
convex case is also supported by empirical evidence, where sampling
strategies tend to outperform basic incremental-gradient methods.

\section{Numerical experiments in seismic inversion}
\label{sec:full-wavef-invers}

A good candidate for the sampling approach we have discussed is the
full waveform inversion problem from exploration geophysics, which we
address using a robust formulation.  The goal is to obtain an estimate
of subsurface properties of the earth using seismic data.  To collect
the data,  explosive charges are detonated just below the surface, and
the energy that reflects back is recorded at the surface by a large
array of geophones.  The resulting data consist of a time-series
collection for thousands of source positions.

The estimate of the medium parameters is based on fitting the recorded
and predicted data.  Typically, the predicted data are generated by
solving a PDE whose coefficients are the features of interest. The
resulting PDE-constrained optimization problem can be formulated in either
the time~\cite{Tarantola1984} or the frequency~\cite{Pratt95} domain.  It
is common practice to use a simple scalar wave equation to predict the
data, effectively assuming that the earth behaves like a fluid---in
this case, sound speed is the parameter we seek.

Raw data are processed to remove any unwanted artifacts; this requires
significant time and effort.  One source of unwanted artifacts in the
data is equipment malfunction.  If some of the receivers are not
working properly, the resulting data can be either zero or
contaminated with an unusual amount of noise. And even if we were to
have a perfect estimate of the sound speed, we still would not expect
to be able to fit our model perfectly to the data. The presence of
these outliers in the data motivates us (and many other authors, e.g.,
\cite{Bube1997,Symes2003,Brossier2010}) to use robust methods for this
application.  We compare the results of robust Student's t-based
inversion to those obtained using least-squares and Huber robust
penalties, and we compare the performance of deterministic,
incremental-gradient, and sampling methods in this setting.

\subsection{Modelling and gradient computation for full waveform
  inversion}

The forward model for frequency-domain acoustic FWI, for a single
source function $q$, assumes that wave propagation in the earth is described by
the scalar Helmholtz equation
\begin{equation*}
  A_{\omega}(x)u = [\omega^2 x + \nabla^2]u = q,
\end{equation*}
where $\omega$ is the angular frequency, $x$ is the squared-slowness
(seconds/meter)$^2$, and $u$ represents the wavefield.  The
discretization of the Helmholtz operator includes absorbing boundary
conditions, so that $A_{\omega}(x)$ and $u$ are complex-valued.  The
data are measurements of the wavefield obtained at the receiver
locations $d = Pu$. The forward modelling operator $F(x)$ is then
given by
\begin{equation*}
  F(x) = PA^{-1}(x),
\end{equation*}
where $A$ is a sparse block-diagonal matrix, with blocks $A_{\omega}$
indexed by the frequencies $\omega$. Multiple sources $q_i$ are
typically modeled as discretized delta functions with a
frequency-dependent weight.  The resulting data are then modeled by
the equation $d_i = F(x)q_i$, and the corresponding residual equals
$r_i(x) = d_i - F(x)q_i$ (cf.~\eqref{eq:2}).

For a given loss function $\rho$, the misfit function and its
gradient are  defined as
\begin{equation*}
 \phi(x)=\sum_{i=1}^{m}\rho(r_i(x))
 \text{and}
 \nabla\phi(x)=\sum_{i=1}^m \nabla F(x,q_i)^*\nabla\rho(r_i(x)),
\end{equation*}
where $\nabla F(x,q_i)$ is the Jacobian of $F(x)q_i$.  The action of the adjoint
of the Jacobian on a vector $y$ can be efficiently computed via the
adjoint-state method \cite{Tarantola1984} as follows:
\begin{equation*}
\nabla F(x, q_i)^*y = G(x,u_i)^*v_i,
\end{equation*}
where $G(x,u_i)$ is the (sparse) Jacobian of $A(x)u_i$ with respect to
$x$, and $u_i$ and $v_i$ are solutions of the linear systems
\[
A(x)u_i = q_i \text{and} A(x)^*v_i = Py.
\]

The Huber penalty function for a vector $r$ is
\begin{equation*}
 \rho(r) = \sum_i \zeta_i,
 \text{where}
 \zeta_i = \begin{cases}
      r_i^2/{2\mu} & \hbox{if $|r_i|\leq \mu$}
    \\|r_i|-\mu/2  & \hbox{otherwise.}
  \end{cases}
\end{equation*}
The Student's t penalty function~\eqref{eq:19} for a vector $r$ is
defined by
\[
 \rho(r)=\sum_i\log(1+r_i^2/\nu).
\]

\subsection{Experimental setup and results}

\begin{figure}[t]
\centering
\includegraphics[width=.5\linewidth]{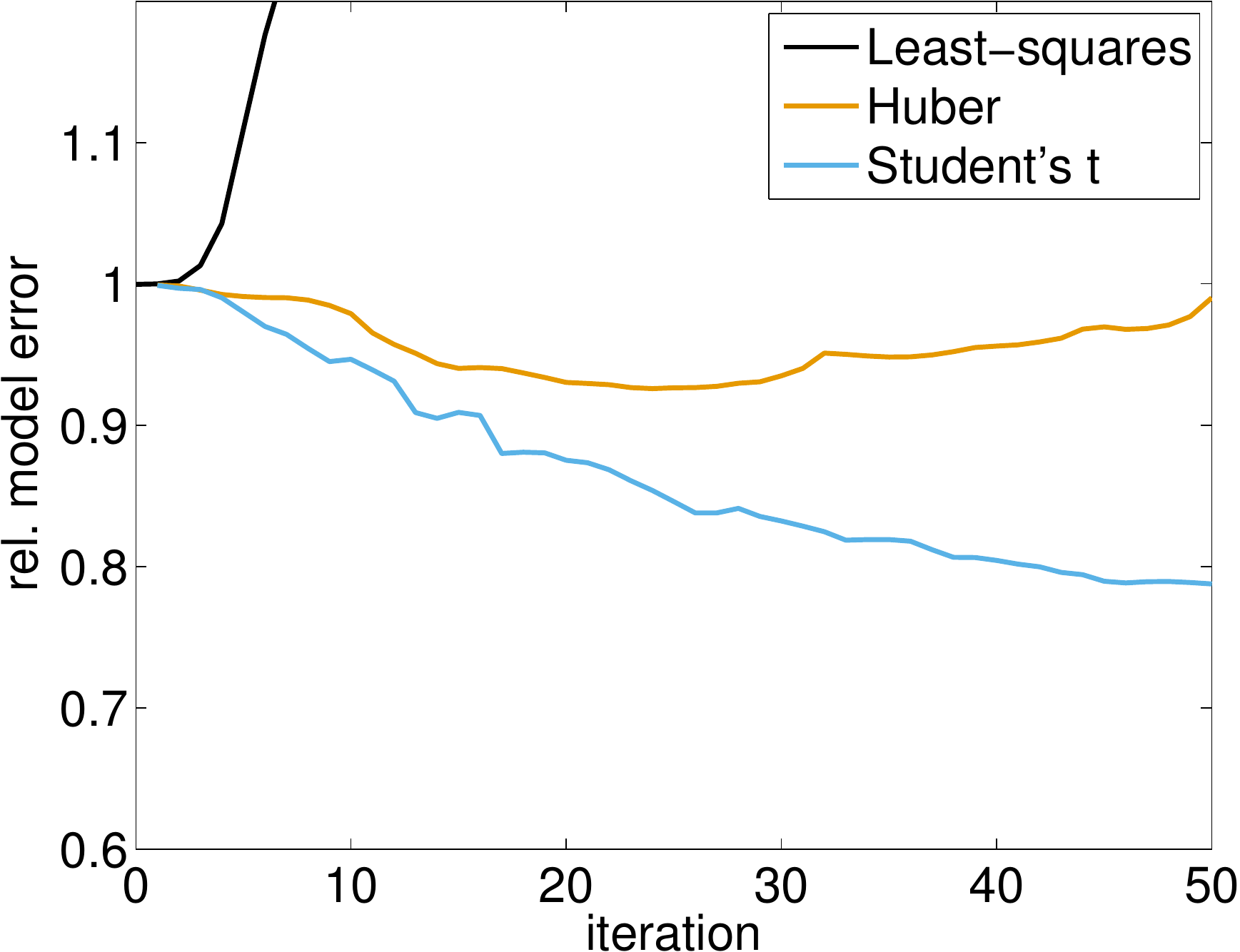}
\caption{Relative error between the true and reconstructed models for
  least-squares, Huber, and Student t penalties. In the least-squares
  case, the model error is not reduced at all. Slightly better results
  are obtained with the Huber penalty, although the model error starts
  to increase after about 20 iterations.  The Students t penalty gives
  the best result.}
\label{fig:fwimodelerror}
\end{figure}

For the seismic velocity model $x^*\in \mathbb{R}^{60501}$ on a
201-by-301 grid depicted in
Figure~\ref{fig:fwiresids}\subref{fig:fwiresids_true}, observed data
$d$ (a complex-valued vector of length 272,706) are generated using 6
frequencies, 151 point sources, and 301 receivers located at the
surface.  To simulate a scenario in which half of the receivers at
unknown locations have failed, we multiply the data with a mask that
zeroes out 50\% of the data at random locations. We emphasize that the
model was blind to this corruption, and so we could have equivalently
added a large perturbation to the data, as was done for example
in~\cite{AravkinLH:2011a}.  The resulting data thus differ from the
prediction $F(x^*)$ given by the true solution $\xstar$. A spike in
the histogram of the residuals $r_i(\xstar)$ evaluated at the true
solution $\xstar$, shown in
Figure~\ref{fig:fwiresids}\subref{fig:fwiresids_true}, shows these
outliers. The noise does not fit well with any simple prior
distribution that one might like to use.
We solve the resulting optimization problem with the least-squares,
Huber, and Student t- penalties using a limited-memory BFGS
method. Figure~\ref{fig:fwimodelerror} tracks across iterations the
relative model error $\|x_k - x^*\|/\|x^*\|$ for all three
approaches. Histograms of the residuals after 50 iterations are
plotted in Figures~\ref{fig:fwiresids}(c)--(e). The residuals for the
least-squares and Huber approaches resemble Gaussian and Laplace
distributions respectively. This fits well with the prior assumption
on the noise, but does not fit the true residual at all. The residual
for the Student's t approach does \emph{not} resemble the prior
distribution at all. The slowly increasing penalty function allows for
enough freedom to let the residual evolve into the true distribution.
\begin{figure}[t]
\centering
\begin{tabular}{cc}
\includegraphics[scale=.3]{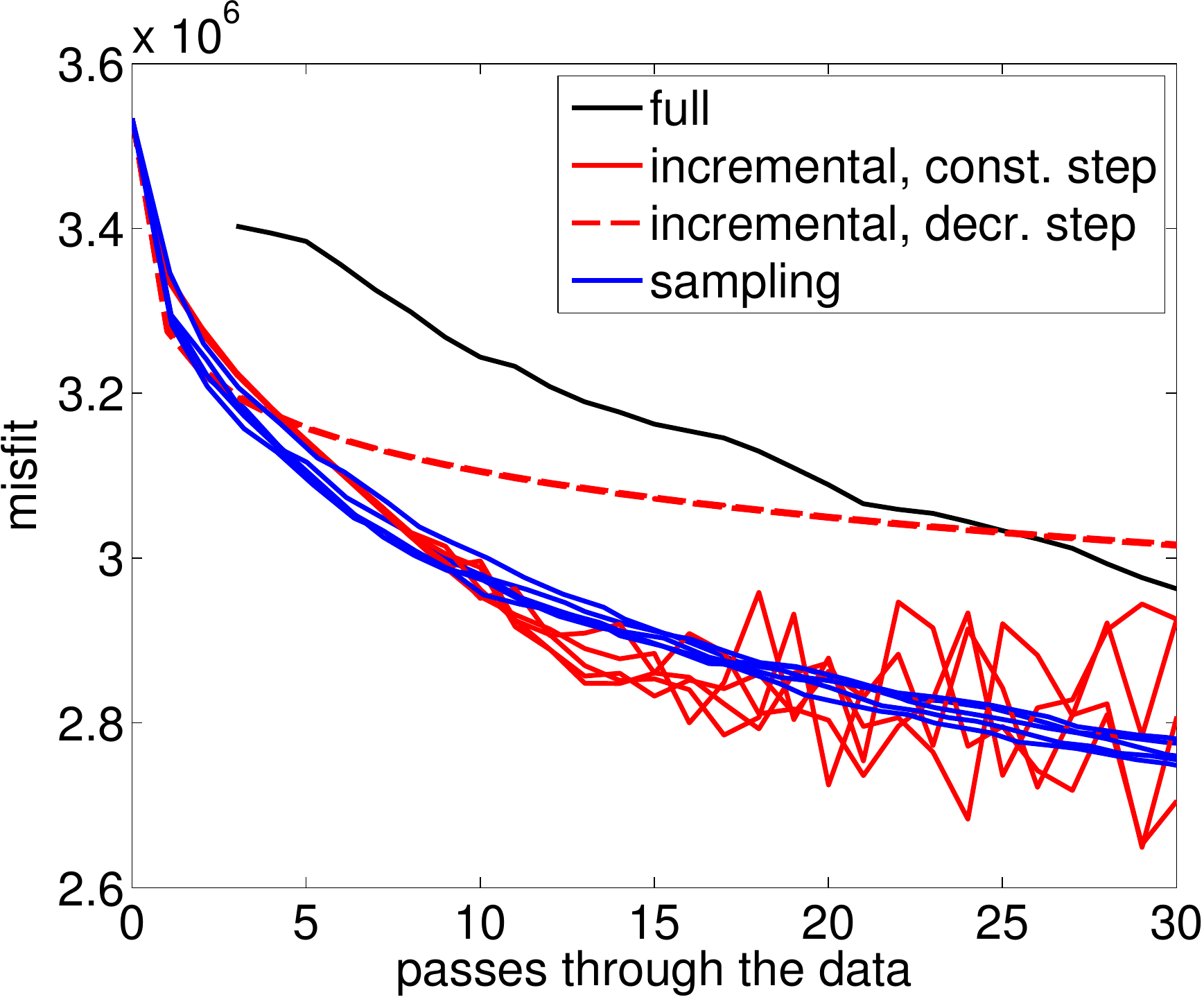}&
\includegraphics[scale=.3]{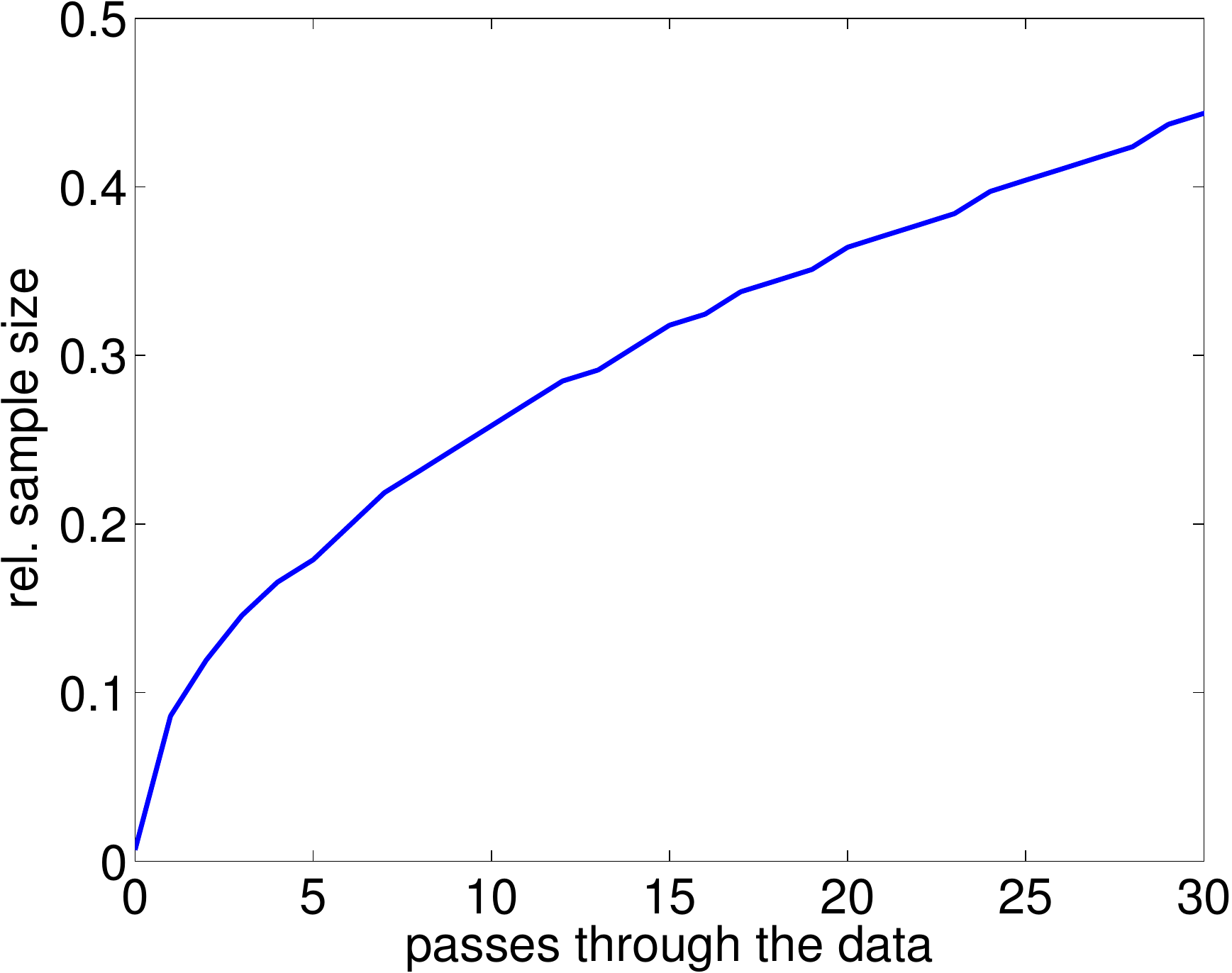}\\
{\small (a)}&
{\small (b)}\\
\end{tabular}
\caption{(a) Convergence of different optimization strategies on the
  Students t penalty: Limited-memory BFGS using the full gradient (``full''),
  incremental gradient with constant and decreasing step
  sizes, and the sampling approach.  Different lines of the same color
  indicate independent runs with different random number streams.  (b)
The evolution of the amount of data used by the sampling method.%
}
\label{fig:fwiconvergence}
\end{figure}

Next, we compare the performance of the incremental-gradient
(\S\ref{sec:incremental-grad}) and sampling (\S\ref{sec:sampling-opt})
algorithms against the full-gradient method.  For the
incremental-gradient algorithm~\eqref{eq:4}, at each iteration we
randomly choose $i$ uniformly over the set $\Set{1,2,\ldots,m}$, and use
either a fixed stepsize $\alpha_k \equiv \alpha$ or a decreasing
stepsize $\alpha_k = \alpha/\lfloor k/m\rfloor$. The sampling method
is implemented via the iteration
\begin{equation*}
 x_{k+1} = x_k - \alpha_k p\k,
\end{equation*}
where $p\k$ satisfies~\eqref{eq:16}, and $H\k$ is a limited-memory
BFGS Hessian approximation. The quasi-Newton Hessian $H_k$ is updated
using the pairs $(\Delta x\k,\Delta g\k)$, where
\[
\Delta x\k := x\kp1 - x\k
\text{and}
\Delta g\k := g\kp1 - g\k;
\]
the limited-memory Hessian is based on a history of length 4. Nocedal
and Wright\cite[\S7.2]{Noce} describe the recursive procedure for
updating $H\k$. The batch size is increased at each iteration by only
a single element, i.e.,
\[
s\kp1 = \min\{\, m,\,s\k + 1\,\}.
\]
The members of the batch are redrawn at every iteration, and we use an
Armijo backtracking linesearch based on the sampled function
$(1/s\k)\sum_{i\in\sample_k}\phi_i(x)$.

The convergence plots for several runs of the sampling method and the
stochastic gradient method with $\alpha = 10^{-6}$ are shown in
Figure~\ref{fig:fwiconvergence}(a).
Figure~\ref{fig:fwiconvergence}(b) plots the evolution of the amounts
of data sampled.

\section{Discussion and conclusions}
\label{sec:disc-concl}

The numerical experiments we have conducted using the Student's t-penalty are
encouraging, and indicate that this approach can overcome some of the
limitations of convex robust penalties such as the Huber norm. Unlike the
least-squares and Huber penalties, the Student t-penalty does not
force the residual into a shape prescribed by the corresponding
distribution. The sampling method successfully combines the
steady convergence rate of the full-gradient method with the
inexpensive iterations provided by the incremental-gradient method.

The convergence analysis of the sampling method, based on
Theorem~\ref{th:g-w-error}, relies on bounding the second moment of
the error in the gradient, and hence the variance of the sample
average (see~\eqref{eq:var-sample-avg}). The bound on the
second-moment arises because of our reliance on the concept of an
\emph{expected} distance to optimality
$\expval[\norm{x\k-\xstar}^2]$. However, other probabilistic measures
of distance to optimality may be more appropriate; this would
influence our criteria for bounding the error in the gradient. For
example, Avron and Toledo~\cite{AvronToledo:2011} measure the quality
of a sample average using an ``epsilon-delta'' argument that provides
a bound on the sample size needed to achieve a particular accuracy
$\epsilon$ with probability $1-\delta$.

Other refinements are possible. For example, van den Doel and
Ascher~\cite{AschervdDoel:2011} advocate an adaptive approach for
increasing the sample size, and Byrd et al.~\cite{byrd:2011} use a
sample average-approximation of the
Hessian.%
\bibliographystyle{siam}
\bibliography{seg2011}

\end{document}